\documentclass[prd,twocolumn,nofootinbib]{revtex4-1}
\usepackage{amsfonts}
\usepackage{amsmath}
\usepackage{amssymb}
\usepackage{bm}
\usepackage{dcolumn}
\usepackage[dvips]{graphicx}
\usepackage{graphics}
\usepackage[normalem]{ulem}
\usepackage{latexsym}
\usepackage{rotating}
\usepackage[colorlinks=true]{hyperref}
\usepackage{xspace} 
\usepackage[usenames]{color}
\usepackage[dvipsnames]{xcolor}
\usepackage{mathrsfs}
\usepackage{multirow}
\usepackage{pifont}
\usepackage{enumitem}
\usepackage{color}
\usepackage{url}
\usepackage{appendix}
\usepackage[utf8]{inputenc}
\usepackage{comment}
\definecolor {darkgreen}{rgb}{0.2,0.7,0.2}
\definecolor{purple}{rgb}{0.5,0,0.5}

\newcommand\be{\begin{equation}}
\newcommand\ba{\begin{eqnarray}}
\newcommand\ee{\end{equation}}
\newcommand\ea{\end{eqnarray}}
\newcommand\bw{\begin{widetext}}
\newcommand\ew{\end{widetext}}
\newcommand{\lb}{\left(}
\newcommand{\rb}{\right)}

\newcommand{\EDGB}{{\mbox{\tiny EdGB}}}
\newcommand{\MG}{{\mbox{\tiny MG}}}

\newcommand{\PPE}{{\mbox{\tiny PPE}}}

\newcommand{\ST}{{\mbox{\tiny ST}}}

\newcommand{\GR}{{\mbox{\tiny GR}}}

\begin{document}
\title{Testing Gravity with Gravitational Waves from Binary Black Hole Mergers: \\ Contributions from Amplitude Corrections}

\author{Shammi Tahura}
\affiliation{Department of Physics, University of Virginia, Charlottesville, Virginia 22904, USA.}

\author{Kent Yagi}
\affiliation{Department of Physics, University of Virginia, Charlottesville, Virginia 22904, USA.}

\author{Zack Carson}
\affiliation{Department of Physics, University of Virginia, Charlottesville, Virginia 22904, USA.}

\begin{abstract}
The detection of gravitational waves has offered us the opportunity to explore the dynamical and strong-field regime of gravity.
Because matched filtering is more sensitive to variations in the gravitational waveform phase than the amplitude, many tests of gravity with gravitational waves have been carried out using only the former. Such studies cannot probe the non-Einsteinian effects that may enter only in the amplitude. Besides, if not accommodated in the waveform template, a non-Einsteinian effect in the amplitude may induce systematic errors on other parameters such as the luminosity distance.
In this paper, we derive constraints on a few modified theories of gravity (Einstein-dilaton-Gauss-Bonnet gravity, scalar-tensor theories, and varying-$G$ theories), incorporating both phase and amplitude corrections. We follow the model-independent approach of the parametrized post-Einsteinian formalism.
We perform Fisher analyses with Monte-Carlo simulations using the LIGO/Virgo posterior samples.
We find that the contributions from amplitude corrections can be comparable to the ones from the phase corrections in case of massive binaries like GW150914.
Also, constraints derived by incorporating both phase and amplitude corrections differ from the ones with phase corrections only by 4\% at most, which supports many of the previous studies that only considered corrections in the phase.
We further derive reliable constraints on the time-evolution of a scalar field in a scalar-tensor theory for the first time with gravitational waves.
\end{abstract}

\date{\today}

\maketitle


\section{Introduction}
So far, general relativity (GR) is the most successful theory of gravitation. This century-old theory which exquisitely describes gravity as the curvature of spacetime has passed numerous tests with high precision~\cite{Will:2014kxa}. Nonetheless, GR is not expected to be a complete description of gravity. The inconsistencies in  galaxy rotation curves~\cite{article,Bosma:1981zz,Begeman:1991iy,Rubin:1970zza,Rubin:1980zd,1973ApJ...186..467O,Ostriker:1993fr} and the accelerated expansion of the universe~\cite{Abbott:1988nx,Copeland:2006wr,Perlmutter:1998np,Riess:1998cb,Riess:2004nr,RevModPhys.61.1,vanAlbada:1984js,WEINBERG201387}  are difficult to explain within the formulation of GR without introducing dark matter and dark energy. Moreover, a new theory is required to reconcile quantum mechanics with classical gravity~\cite{Adler:2010wf,Ng:2003jk}. Hence, one needs to continue testing GR through various experiments and observations. Gravitational wave (GW) observations are one of the most recent additions to this venture~\cite{TheLIGOScientific:2016src,Yunes:2016jcc,LIGOScientific:2019fpa,Monitor:2017mdv,Abbott:2018lct} which have enabled us to probe the formerly inaccessible strong, highly non-linear and dynamical regime of gravity. Since the strong-field regime is precisely the place to look for evidence of beyond-GR phenomena due to quantum gravity corrections~\cite{Giddings:2019ujs,Carballo-Rubio:2018jzw}, it is important to extract as much physics as possible from the available GW data.


One can adopt either a model-independent or a theory-specific method for testing gravity, although the former is more efficient if one wishes to achieve constraints on multiple theories with GW observations. One of the first works on the theory-agnostic approach was taken in~\cite{Arun:2006yw,Arun:2006hn,Mishra:2010tp} where each post-Newtonian (PN) term in the GR waveform phase were treated independent and the authors proposed to study the consistency among them. One drawback of such an approach is that it cannot capture the non-GR effects entering at PN orders that are absent in GR (like $-1$PN order common in scalar-tensor theories). To overcome this, Yunes and Pretorius proposed a new framework called \emph{parametrized post-Einsteinian} (PPE) formalism by introducing generic corrections at any PN order to both the phase and the amplitude~\cite{Yunes:2009ke,Chatziioannou:2012rf}.
A theory-agnostic data analysis pipeline named TIGER has been developed~\cite{Agathos:2013upa,Meidam:2014jpa} and the LIGO and VIRGO Scientific Collaboration (LVC) recently employed the generalized IMRPhenom (gIMR) waveform model which has a one-to-one mapping with the PPE formalism in the inspiral part of the waveform phase. With such waveforms, tests of gravity with the GW phase have been carried out in~\cite{Monitor:2017mdv,TheLIGOScientific:2016pea,Yunes:2016jcc,Abbott:2017vtc,LIGOScientific:2019fpa,TheLIGOScientific:2016src}. 

Many of the previous studies on tests of GR with GWs focused only on the phase corrections, though scenarios where amplitude corrections bear importance  are not uncommon. In some parity-violating theories, one of the circularly-polarized modes is amplified while the other one is suppressed, an effect called amplitude birefrigence~\cite{Alexander:2007kv,Yunes:2008bu,Yunes:2010yf,Yagi:2017zhb}. Such an effect enters only in the GW amplitude of circularly-polarized modes. Probing amplitude corrections is also important in constraining gravitational theories with GW stochastic backgrounds~\cite{Maselli:2016ekw}. Furthermore, theories with flat extra dimensions~\cite{Cardoso:2002pa}, Horndeski gravity~\cite{Saltas:2014dha}, and $f(R)$ gravity~\cite{Hwang:1996xh} may predict amplitude damping that scales with the cosmological distance. Such phenomena have been studied in Ref.~\cite{Nishizawa:2017nef}  in terms of a generalized GW propagation framework. Possible bounds on the PPE amplitude parameters at various PN orders were studied in~\cite{Cornish:2011ys} while both amplitude and phase corrections were included in~\cite{Arun:2012hf} for generic theories with scalar dipole radiation.

We here study how much impact the amplitude corrections may bring to tests of GR with GWs and provide justifications for previous studies that only considered the phase corrections.
PPE amplitude corrections due to generation mechanisms in various example theories have been derived analytically in Ref.~\cite{Tahura:2018zuq}. We compute the constraints on some of those theories from both the phase and amplitude, focusing on leading PN corrections to the tensorial modes only. We choose theories where the leading correction enters at a negative PN order, and the sensitivities of black holes (BHs) are known. Such criteria lead us to choose Einstein-dilaton-Gauss-Bonnet (EdGB) gravity, scalar-tensor theories, and varying-$G$ theories. We carry out Fisher analyses with Monte-Carlo simulations utilizing the parameter posterior samples of GW151226 and GW150914 released by LVC~\cite{ligo:sample}\footnote{\bf{We choose GW151226 and GW150914 as representatives of low-mass and massive binaries respectively, following \cite{Yunes:2016jcc}.}}. Such analyses with actual posterior samples produce more reliable results compared to the ones with sky-averaged waveforms. In fact, when implementing such samples, we can determine the credibility of the small coupling approximation in scalar-tensor theories, which allows us to place reliable bounds on the time-evolution of the scalar field from GW observations for the first time.

{
\newcommand{\minitab}[2][l]{\begin{tabular}{#1}#2\end{tabular}}
\renewcommand{\arraystretch}{2.}
\begingroup 
\begin{table*}[htb]
\begin{centering}
\begin{tabular}{c|c|c|c|c|c|c|c}
\hline
\hline
\noalign{\smallskip}
\multirow{3}{*}{Theories}&\multirow{3}{*}{Repr. Parameter}&\multicolumn{6}{c}{Constraints} \\ \cline{3-8}
&&\multicolumn{3}{c|}{GW150914}&\multicolumn{3}{c}{GW151226}  \\ \cline{3-8}
&&Phase&Amplitude&Combined&Phase&Amplitude&Combined\\ \hline
\multirow{2}{*}{EdGB~\cite{Yagi:2011xp}}&$\sqrt{|\bar{\alpha}_{\EDGB}|}$ [km]&(50.5)&(76.3)&(51.5)&4.32&10.5&4.32\\
&$\zeta_{\EDGB}$&3.62&32.4&3.91&0.0207&0.709&0.0207\\ \hline
\multirow{2}{*}{Scalar-Tensor~\cite{Scharre:2001hn,Berti:2004bd}}&$|\dot{\phi}|$ [$10^4/\mathrm{sec}$]&(3.64)&(7.30)&(3.77)&1.09&(5.60)&1.09\\
&$|m_1\dot{\phi}|$&6.87&16.4&7.15&0.688&3.66&0.688\\ \hline
Varying-$G$~\cite{Tahura:2018zuq,Yunes:2009bv}&$|\dot{G_0}/G_0|$ [$10^6/\mathrm{yr}$]&7.30&137&7.18&0.0224&0.382&0.0220\\ 
\noalign{\smallskip}
\hline
\hline
\end{tabular}
\end{centering}
\caption{90\% credible constraints on representative parameters of various modified theories of gravity from GW150914 and GW151226. For each of the GW events, ``phase'' and ``amplitude'' correspond to the cases where we include non-GR corrections only to the GW phase and amplitude respectively, while ``combined'' is the case where we include both corrections in the waveform and reduce the two constraints to a single one according to Sec.~\ref{data}. $\bar{\alpha}_{\EDGB}$ is the EdGB coupling parameter which is related to the dimensionless coupling by $\zeta_\EDGB\equiv 16 \pi \bar{\alpha}_\EDGB^2/m^4$ with $m$ being the total mass of the binary. $m_1\dot{\phi}$ corresponds to a dimensionless parameter in scalar-tensor theories where $m_1$ is the mass of the primary BH while $\phi$ is the scalar field. The bounds are derived by assuming subdominant non-GR corrections,  which is realized whenever $\zeta_\EDGB \ll 1$ ($m_1\dot{\phi} \ll 1$) in EdGB (scalar-tensor) gravity. Numbers inside brackets mean such criterion is violated and the constraints are unreliable. $G$ is the gravitational constant with the subscript 0 representing the time of coalescence. An overhead dot denotes a derivative with respect to time.}
\label{table:ppE}
\end{table*}
\endgroup
}

We find that the constraints derived from the phase and the amplitude can be comparable in case of massive binary systems like GW150914. Whereas for less massive binaries with a larger number of GW cycles, the phase always yields stronger constraints. Moreover, inclusion of an amplitude correction to the waveform impacts the bound on the phase correction as well since the former can easily be related to the latter provided the dissipative and conservative corrections do not enter at the same order. The amount and direction of such effects vary with the PN order of the corrections. All such constraints in the theories under consideration are summarized in Table~\ref{table:ppE}.

The rest of the paper is organized as follows. Section~\ref{section:ppE} briefly reviews PPE formalism while Sec.~\ref{data} summarizes the data analysis techniques. Section~\ref{sec:massive} is devoted to justifying our formalism against the one by LVC in massive gravity~\cite{TheLIGOScientific:2016src} while we derive constraints on EdGB, scalar-tensor, and varying-$G$ theories in Secs.~\ref{EdGB}-~\ref{vaying-G}. Section~\ref{conclusion} presents a summary of our work while discussing the effects of an amplitude correction on that of phase. Appendix~\ref{Appendix} compares the PhenomB and PhenomD waveforms for constraining PPE parameters.

\section{Methodology}\label{section:method}

In this section, we explain how we perform our analysis. We first explain the PPE formalism and the non-GR waveform template. We then describe the Fisher analysis and how we construct probability distributions of non-GR parameters.

\subsection{PPE Waveform}\label{section:ppE}
We begin by reviewing the PPE formalism briefly. PPE gravitational waveform for a compact binary inspiral in the frequency domain is given by~\cite{Yunes:2009ke}
\begin{equation}\label{eq:2a}
\tilde{h}(f)=\tilde{h}_{\GR}(1+\alpha_{\PPE}\, u^a)e^{i\delta\Psi}\,,
\end{equation}
where $\tilde{h}_{\GR}$ is the gravitational waveform in GR. $\alpha_{\PPE}\, u^a$ is a correction to the GW amplitude with $u\equiv(\pi \mathcal{M} f)^{1/3}$, $\mathcal{M}\equiv(m_1m_2)^{3/5}/(m_1+m_2)^{1/5}$ is the chirp mass with component masses $m_1$ and $m_2$, and $f$ is the frequency of the GW. The constant $\alpha_{\PPE}$ controls the overall magnitude of the correction, while the index $a$ specifies at which PN order the correction enters. One can write the non-GR phase correction $\delta\Psi$ in a similar manner as that of the amplitude as
\begin{equation}\label{eq:2b}
\delta\Psi=\beta_{\PPE} u^b\,.
\end{equation}
Together $\left(\alpha_{\PPE},a\right)$ and $\left(\beta_{\PPE},b\right)$ are called the PPE parameters.

PPE modifications in Eq.~\eqref{eq:2a} can enter through the non-GR corrections to the binding energy and the GW luminosity~\cite{Yunes:2009ke,Chatziioannou:2012rf}, or alternatively to the frequency evolution and the Kepler's law~\cite{Tahura:2018zuq}. We will follow the latter approach and write the modified Kepler's law as
\begin{equation}
 \label{eq:2c}
 r=r_{\GR}(1+\gamma_r u^{c_r})\,,
 \end{equation}
and the frequency evolution as
\begin{equation}\label{eq:2d}
\dot{f}=\dot{f}_{\GR}\left(1+\gamma_{\dot{f}}u^{c_{\dot{f}}}\right)\,.
\end{equation}
Here, $\lb\gamma_r,c_r\rb$ and $\lb\gamma_{\dot{f}},c_{\dot{f}}\rb$ parametrize the non-GR corrections to the binary separation $r$ and the frequency evolution $\dot{f}$ respectively. To leading PN order, the GR contribution is given by~\cite{cutlerflanagan,Blanchet:1995ez}
\begin{equation}
r_{\GR}=\left(\frac{m}{\Omega^2}\right)^{1/3}\,, \quad
\dot{f}_{\GR}=\frac{96}{5}\pi^{8/3}\mathcal{M}^{5/3}f^{11/3}\,,
\end{equation}
where $m$ represents the total mass of the binary and $\Omega=\pi f$ is the orbital angular frequency.

Utilizing the stationary phase approximation~\cite{PhysRevD.62.084036,Yunes:2009yz} and the quadrupole formula for the metric perturbation~\cite{Blanchet:2002av}, one can easily derive the amplitude and phase of the dominant quadrupolar mode in Fourier space from Eqs.~\eqref{eq:2c} and~\eqref{eq:2d} as
\begin{equation}\label{eq:amp}
\tilde{\mathcal{A}}(f)=\tilde{\mathcal{A}}_{\GR} \left(1+2\gamma_ru^{c_r}-\frac{1}{2}\gamma_{\dot{f}}u^{c_{\dot{f}}}\right)\,,
\end{equation}
and
\be
\label{eq:Psi}
\Psi = \Psi_\GR  -\frac{15 \text{$\gamma_{\dot{f}} $}}{16 (\text{$c_{\dot{f}}$}-8) (\text{$c_{\dot{f}}$}-5)} u^{c_{\dot{f}}-5}\,,
\ee
respectively. Eq.~\eqref{eq:Psi} is already in the PPE format, while  Eq.~\eqref{eq:amp} can be reduced to such a form by keeping only the dominant correction\footnote{A detailed derivation can be found in Ref.~\cite{Tahura:2018zuq}}.

In fact, the PPE phase and the amplitude parameters may be related as follows. If the dissipative correction (correction entering in the GW luminosity) dominates over the conservative one (correction entering in the binding energy and Kepler's law), we find
\begin{equation}\label{eq:2w2}
\alpha_{\PPE} = \frac{8}{15} (a-8)(a-5) \, \beta_{\PPE}\,,
\end{equation}
while for the conservative-dominated case, we obtain
\begin{equation}\label{eq:2w3}
\alpha_{\PPE} =\frac{8}{15} \frac{(8-a)(5-a)(a^2-4a-6)}{a^2-2a-6} \beta_{\PPE}\,.
\end{equation}
On the other hand, when the aforementioned corrections enter at the same PN order, no direct relation between $\alpha_{\PPE}$ and $\beta_{\PPE}$ exists. The exponents $a$ and $b$ in the correction terms are related by the following equation which is valid for all three cases:
\begin{equation}
b=a-5\,.
\end{equation}

The above formalism needs to be slightly modified for theories containing time-varying gravitational constants. Variations in the gravitational constants cause the masses  of the binary components to vary as well~\cite{PhysRevLett.65.953}, and one needs to take this into account when deriving the PPE parameters~\cite{Tahura:2018zuq}.
\subsection{Data Analysis Formalism}\label{data}

We adopt a Fisher analysis~\cite{Cutler:1994ys} to estimate the statistical errors of the non-GR parameters in various theories. Such an analysis is valid for GW events with sufficiently large signal-to-noise (SNR) ratios. We make the assumptions that the detector noise is Gaussian and stationary. Let us write the detector output as
\be
s(t)=h(t)+n(t)\,,
\ee
where $h(t)$ and $n(t)$ are the GW signal and the noise respectively. Let us also define the inner product of two quantities $A(t)$ and $B(t)$ as
 \be
 \left(A|B\right)=4\Re\int_0^\infty df \frac{\tilde{A}^*(f)\tilde{B}(f)}{S_n \left( f \right)}\,.
 \ee
Here $\tilde{A}(f)$ is the Fourier component of $A$, an asterisk ($*$) superscript means the complex conjugate and $S_n\left(f\right)$ is the noise spectral density. With the above definitions, the probability distribution of the noise can be written as
 \be
P\left(n=n_0(t)\right) \propto \text{exp}\left[-\left(n_0|n_0\right)\right]\,,
\ee
and the SNR for a given signal $h(t)$ can be defined as
\be
\rho \equiv\sqrt{\left(h|h\right)}\,.
\ee
Under the assumptions of Gaussian and stationary noise, the posterior probability distribution of binary parameters $\theta^a$ takes the following form:
\be
P\left(\theta^a|s\right) \propto p^{(0)}\left(\theta^a\right) \text{exp}\left[-\frac{1}{2} \Gamma_{ab} \Delta\theta^a\Delta \theta^b \right]\,,
\ee
where $\Delta \theta^a=\hat{\theta}^a-\theta^a$ with $\hat{\theta}^a$ being the maximum likelihood values of $\theta^a$. $p^{(0)}\left(\theta^a\right)$ gives the probability distribution of the prior information, which we take to be in a Gaussian form for simplicity. $\Gamma_{ab}$ is called the Fisher information matrix which is defined as
\be
\Gamma_{ab}=\left(\partial_a h|\partial_b h \right)\,,
\ee
where $\partial_b\equiv \frac{\partial}{\partial \theta^b}$. One can estimate the root-mean-square of $\Delta \theta^a$ by taking the square root of the diagonal elements of the inverse Fisher matrix $\Sigma^{ab}$: 
\be
\Sigma^{ab}=\left(\tilde{\Gamma}^{-1}\right)^{ab}=\langle\Delta\theta^a\Delta \theta^b\rangle\,,
\ee
where $\tilde{\Gamma}_{ab}$ is defined by
\be
p^{(0)}\left(\theta^a\right) \text{exp}\left[-\frac{1}{2} \Gamma_{ab} \Delta\theta^a\Delta \theta^b \right]=\text{exp}\left[-\frac{1}{2} \tilde{\Gamma}_{ab} \Delta\theta^a\Delta \theta^b \right]\,.
\ee

To save computational time, we use IMRPhenomB waveform. Reference~\cite{Yunes:2016jcc} showed that the difference in constraints on PPE phase parameters between IMRPhenomB and IMRPhenomD waveforms are negligible for propagation mechanisms at any PN order and for generation mechanisms at negative PN orders.
In App.~\ref{Appendix}, we perform a similar comparison for generation mechanism corrections in the amplitude using sky-averaged waveforms and show that the former is at least suitable for constraining generation mechanisms that enter at negative PN orders, which is what we will consider in Sec.~\ref{results}.

We choose the following parameters as our variables for the Fisher analysis:
\ba\label{parameters}
\theta^a\equiv \lb \ln{\mathcal{M}_z}, \ln{\eta}, \chi, \ln{D_L}, \ln{t_0},  \phi_0,\alpha, \delta, \psi, \iota, \theta_\PPE \rb\,, \nonumber \\
\ea
where $\mathcal{M}_z$ is the redshifted chirp mass, $\eta \equiv m_1m_2/(m_1+m_2)^2$ is the symmetric mass ratio and $\chi$ is the effective spin parameter\footnote{The effective spin parameter is defined as $\chi\equiv\lb m_1 \chi_1+m_2\chi_2\rb /\lb m_1+m_2\rb$, where $\chi_A$ with $A=(1,2)$ is the dimensionless spin of the $A$th body.}. $\alpha, \delta, \psi$, and  $\iota$ are the right ascension, declination, polarization and inclination angles respectively in the detector frame. The non-GR parameter is represented by $\theta_\PPE=\alpha_{\PPE}$ or $\beta_{\PPE}$.
We perform a Monte Carlo simulation by using each set of the posterior samples released by LIGO~\cite{ligo:sample} for $(\mathcal{M}_z, \eta, D_L, \chi, \alpha, \delta, \iota)$, while we randomly sample the polarization angle $\psi$ and the coalescence phase $\phi_0$ in $[0,\pi]$ and $[0,2\pi]$ respectively.
We impose prior information such that $-1 \leq \chi \leq 1$, $-\pi \leq (\phi_0, \alpha, \psi) \leq \pi$, and $-\pi/2 \leq ( \delta, \iota) \leq \pi/2$.

We use the detector sensitivity of Advanced LIGO (aLIGO) O1 run~\cite{LIGOScientific:2018mvr}, and we consider the two detectors at Hanford and Livingston. For simplicity, we assume that the Livingston noise spectrum is identical to that of  Hanford~\cite{Yunes:2009yz}. For the Fisher integration, the minimum frequency is taken to be 20 Hz while the maximum frequency is same as the cutoff frequency above which the signal power is negligible~\cite{Ajith:2009bn}.

Now we are going to discuss how we compute the probability distribution of a non-GR parameter from the output of a Fisher analysis with a Monte Carlo simulation. We set the  fiducial value of any non-GR parameter to be zero for our analysis. We perform the following integration numerically to obtain the compound probability density function\footnote{If the distribution of a random variable $y$ depends on a parameter $x$, and if $x$ follows a certain distribution $P(x)$ (called the mixing or latent distribution), the marginal distribution of $y$ is called mixture distribution or compound probability distribution and is given by $P\left(y\right)=\int P\left(y|x\right) P\left(x\right)dx$~\cite{2016arXiv160204060R}.} of any parameter $\xi$~:
\be
\label{eq3:1}
P\lb\xi\rb=\int P\lb \xi|\sigma_{\xi}\rb P\lb \sigma_{\xi}\rb d\sigma_{\xi}\,,
\ee
where $P\lb\xi\rb$ is the marginal (unconditional) probability density function of $\xi$. $P\lb \xi|\sigma_{\xi}\rb \propto \exp[-(\xi-\bar \xi)^2/2\sigma_{\xi}^2]$ is the conditional probability density function of $\xi$ which we assume to be a Gaussian distribution with a mean $\bar \xi$ and a standard deviation $\sigma_\xi$. $P(\sigma_\xi)$ is the probability distribution of $\sigma_\xi$ computed from the Fisher analysis for the entire posterior distribution.

Let us finish this section by explaining how we can utilize both amplitude and phase corrections to derive constraints on some theory. One can include $\alpha_{\PPE}$ or $\beta_{\PPE}$ as variables to the Fisher analysis as in Eq.~\eqref{parameters} and map them to a non-GR parameter of a theory to derive constraints from the phase and amplitude independently. We refer to such constraints as  the ``phase-only'' and ``amplitude-only'' bounds respectively. How can we achieve a single constraint that accommodates both of them? Recall the relations between the PPE parameters in Sec.~\ref{section:ppE}. One can rewrite $\alpha_{\PPE}$ in the waveform in terms of $\beta_{\PPE}$ according to Eqs.~\eqref{eq:2w2} or~\eqref{eq:2w3} and eliminate the former variable from the analysis. 
We refer to such constraints as the ``phase \& amplitude combined'' bounds\footnote{Alternatively, one can rewrite the PPE corrections in the phase and the amplitude  in terms of non-GR parameters of a theory. Performing Fisher analyses with such parameters as variables lead to similar constraints as the `phase \& amplitude combined'' bounds, although such an approach is not theory-agnostic.}.


\section{Results in Example Theories}\label{results}


We now apply our analysis to some example theories. We begin by studying massive gravity that yields corrections in the phase through propagation mechanisms. We compare bounds from the Fisher analysis to those from LVC's Bayesian analysis to justify the former. We next study EdGB gravity, scalar-tensor theories and varying-$G$ theories, which achieve the corrections through generation mechanisms entering at negative PN orders.
Bounds on these theories are summarized in Table~\ref{table:ppE}.


\subsection{Validation of the Fisher Analysis: Massive Gravity}
\label{sec:massive}

The idea of introducing the mass to gravitons is rather old~\cite{Fierz:1939ix}, and many attempts have been made to construct a feasible theory that allows one to do so~\cite{deRham:2014zqa}.  Such a theory may arise in higher-dimensional setups~\cite{Hinterbichler:2011tt} and has the potential to solve the cosmic acceleration problem~\cite{deRham:2014zqa}. Although gravitons with non-vanishing masses may have additional polarizations as well~\cite{dePaula:2004bc}, we here restrict our attention to the non-GR effects on the tensor modes due to a massive dispersion relation.


We will focus on the non-GR corrections specifically to the GW phase. Thus, the purpose of this section is simply to compare our Fisher analysis with the Bayesian one performed by the LVC. Gravitons with a non-vanishing mass travel at a speed smaller than the speed of light and the non-GR effects accumulate over the distance. Modified dispersion relation for such gravitons is given by $ E^2=p^2c^2+m_g^2c^4$, where $m_g$ is the mass of the graviton while $E$ and $p$ are the energy and the momentum respectively. The PPE phase parameters are~\cite{Will:1997bb}
\be
\beta_\MG=\frac{\pi^2}{\lambda_g^2}\frac{\mathcal{M}}{1+z}D\,, \qquad \qquad b=3\,,
\ee
where
\be
D=\frac{z}{H_0\sqrt{\Omega_M+\Omega_{\Lambda}}}\left[1-\frac{z}{4}\left(\frac{3\Omega_M}{\Omega_M+\Omega_{\Lambda}} \right )+\mathcal{O}(z^2)\right]\,.
\ee
Here, $\Omega_M$ and $\Omega_\Lambda$ are the energy density of matter and dark energy respectively. $H_0$ is the Hubble constant while $z$ is the redshift of the source. $\lambda_g$ is the Compton wavelength of the graviton that is related to $m_g$ as $\lambda_g\equiv h/\lb m_gc\rb$, where $h$ is Planck's constant.

\begin{figure}[htb]
\includegraphics[width=8.5cm]{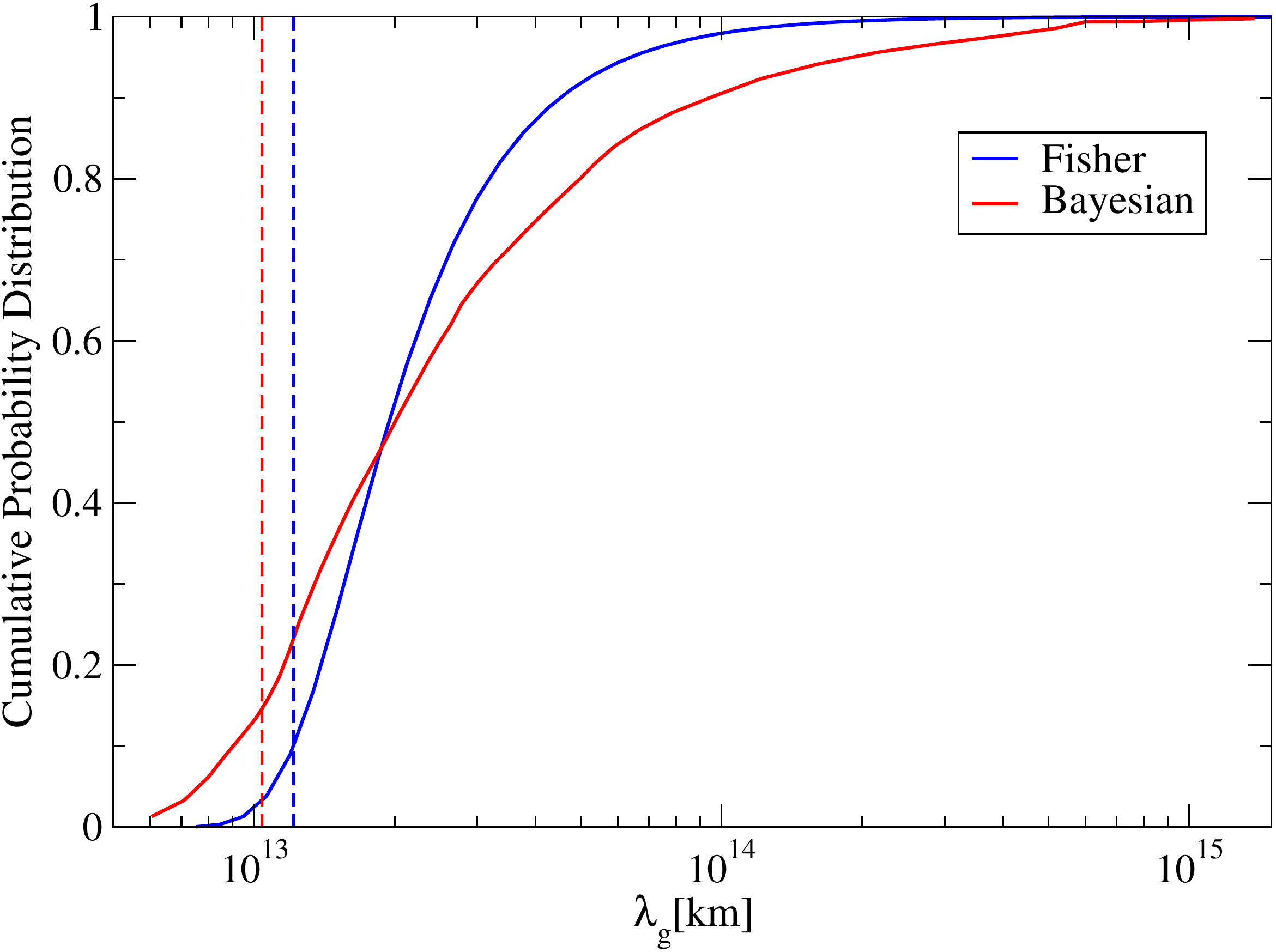}
\caption{The cumulative probability distribution of the graviton Compton wavelength from GW150914. We show the ones obtained from a Fisher analysis with Monte Carlo simulations (blue solid) and from a Bayesian analysis by the LVC (red solid). Each of the vertical dashed lines corresponds to the lower bound of the distribution of the same color with 90\% confidence. Observe how the two different analyses give similar bounds.}
\label{fig:graviton}
\end{figure}

We compute the probability distribution of $\lambda_g$ from GW150914 according to the procedure outlined in Sec.~\ref{section:method} and compare with the one obtained by the LVC~\cite{TheLIGOScientific:2016src} (Fig.~\ref{fig:graviton}). The Fisher analysis with Monte Carlo simulations yields $\lambda_g<1.2\times 10^{13}$ km at 90\% CL, which is in a good agreement with the LVC bound of $1.0\times10^{13}$ km and thus shows the validity of the former. The difference in the two cumulative distributions of $\lambda_g$ presented in Fig.~\ref{fig:graviton} can be attributed to the fact that the LVC used a more accurate Bayesian analysis and imposed a uniform prior on the graviton mass.
The GW bound has recently been updated by combining multiple events~\cite{LIGOScientific:2019fpa}. The new bound is stronger than binary pulsar constraints~\cite{Finn:2001qi,Miao:2019nhf} but slightly weaker than the updated solar system bounds~\cite{Will:2018gku}. The bound is also weaker than the ones from the observations of galactic clusters~\cite{Goldhaber:1974wg,Gupta:2018hgm,Desai:2017dwg}, gravitational lensing~\cite{Choudhury:2002pu}, and the absence of superradiant instability in supermassive BHs~\cite{Brito:2013wya}.

\subsection{Einstein-Dilaton-Gauss-Bonnet Gravity}\label{EdGB}
EdGB gravity endows one of the simplest high-energy modifications to GR~\cite{Moura:2006pz,Pani:2009wy}. Such a theory is motivated from low-energy effective string theories and also arises as a special case of Horndeski gravity~\cite{Zhang:2017unx,Berti:2015itd}. 
The EdGB action is given by introducing a quadratic-curvature correction (Gauss-Bonnet invariant) to the GR action which is non-minimally coupled to a scalar field (dilaton) with a coupling constant $\bar{\alpha}_\EDGB$~\cite{Kanti:1995vq}\footnote{We use barred quantities for coupling constants in order to distinguish them from the PPE parameters.}.

In EdGB gravity, BHs acquire scalar monopole charges which may generate scalar dipole radiation if they form binaries~\cite{Yagi:2011xp,Sotiriou:2014pfa,Berti:2018cxi,Prabhu:2018aun}. Such radiation leads to an earlier coalescence of BH binaries compared to that of GR and modifies the gravitational waveform with the PPE parameters given by~\cite{Yunes:2016jcc,Yagi:2011xp}
\begin{equation}
 \beta_\EDGB=-\frac{5}{7168}\zeta_\EDGB\frac{(m_1^2\tilde s_2^\EDGB-m_2^2\tilde s_1^\EDGB)^2}{m^4\eta^{18/5}}\,,
 \end{equation}
 with $b=-7$ and 
  \begin{equation}
 \alpha_\EDGB=-\frac{5}{192}\zeta_\EDGB\frac{(m_1^2 \tilde s_2^\EDGB-m_2^2 \tilde s_1^\EDGB)^2}{m^4\eta^{18/5}}\,,
 \end{equation}
 with $a=-2$.  Here, $\zeta_\EDGB\equiv 16 \pi \bar{\alpha}_\EDGB^2/m^4$ is the dimensionless EdGB coupling parameter and $\tilde s_{A}^\EDGB$ are the spin-dependent factors of the BH scalar charges given by $\tilde s_{A}^\EDGB\equiv 2(\sqrt{1-{\chi_A}^2}-1+{\chi_A}^2)/{\chi_A}^2~$~\cite{Berti:2018cxi,Prabhu:2018aun}\footnote{For ordinary stars like neutron stars $\tilde s_A^\EDGB$ are zero~\cite{Yagi:2011xp,Yagi:2015oca}.}.

\begin{figure}[htb]
\includegraphics[width=8.5cm]{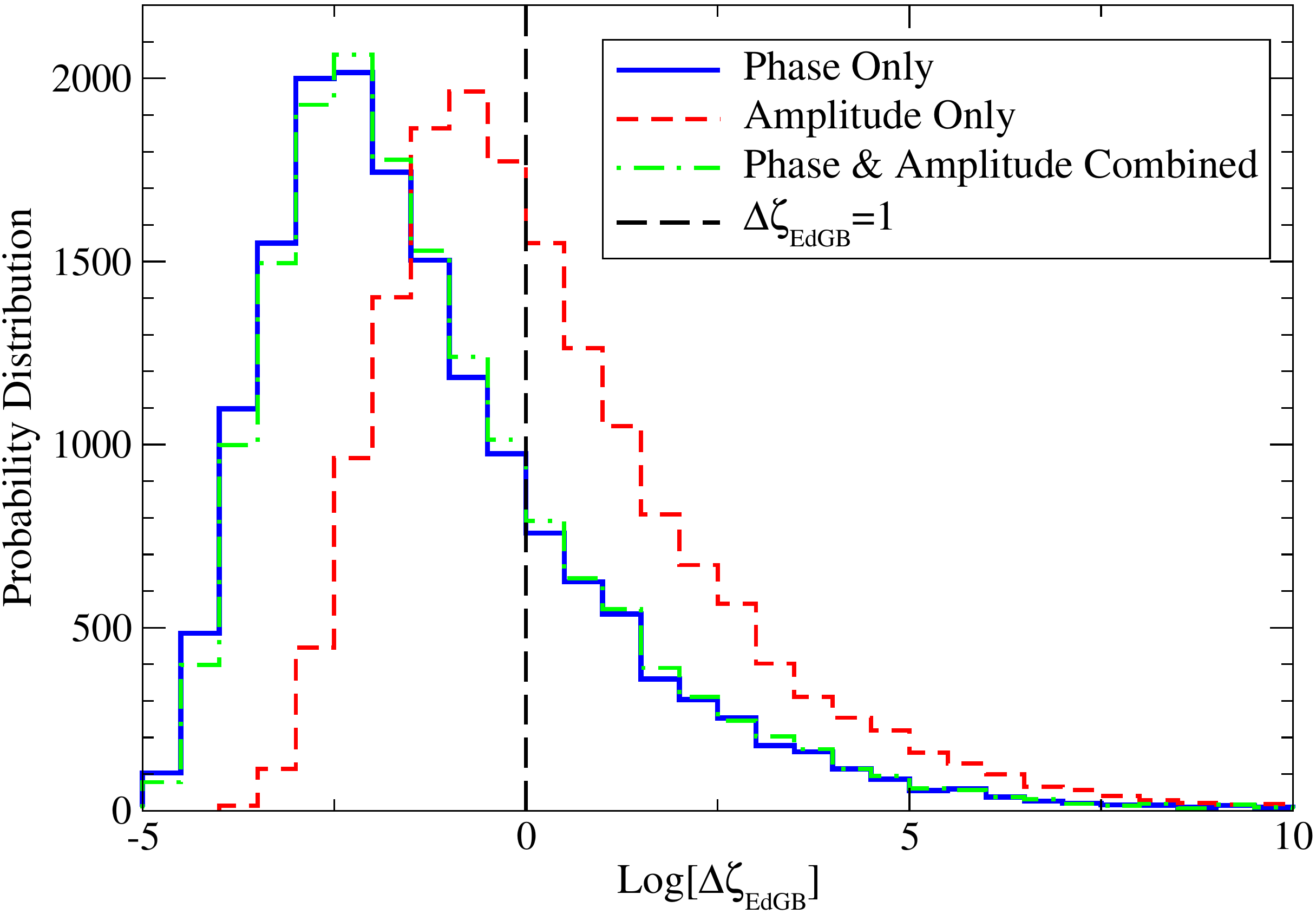}
\caption{Histogram distributions of the 90\% CL bounds on $\zeta_{\EDGB}$ from a Fisher analysis with the phase correction only (blue solid), the amplitude correction only (red dashed) and combining the two corrections (green dotted-dashed). Fiducial values are taken from the posterior samples of GW150914. The samples that lie on the left side of the vertical black dashed line satisfy the small coupling approximation with 90\% CL.}
\label{fig:histogram-edgb}
\end{figure}

We now derive constraints on EdGB gravity from GW150914 and GW151226. First, we estimate how well these events satisfy the small coupling approximation $\zeta_{\EDGB}<1$. To do so, we extract the 90\% CL upper bound $\Delta\zeta_{\EDGB}$ from each sample of the posterior distribution of a particular event. We then create histograms with all the samples (see Fig.~\ref{fig:histogram-edgb}) and calculate the fraction satisfying $\Delta\zeta_{\EDGB}<1$.  For GW150914, 72\% (42\%) of the samples satisfy the small coupling approximation if $\Delta\zeta_{\EDGB}$ is derived from the phase (amplitude) correction only, while 71\% of the posterior distribution satisfies such approximation if the phase and amplitude corrections are combined. A similar analysis with GW151226 gives 98\% and 87\% for the phase and amplitude corrections respectively while combining the two yields almost the same result as that of the phase-only case. Since the fraction of samples satisfying $\zeta_{\EDGB}<1$ is much higher for GW151226 than GW150914 due to a larger number of GW cycles and slower relative velocity of the binary constituents, the former event places more reliable constraints on EdGB gravity compared to the latter one.

\begin{figure}[htb]
\includegraphics[width=8.5cm]{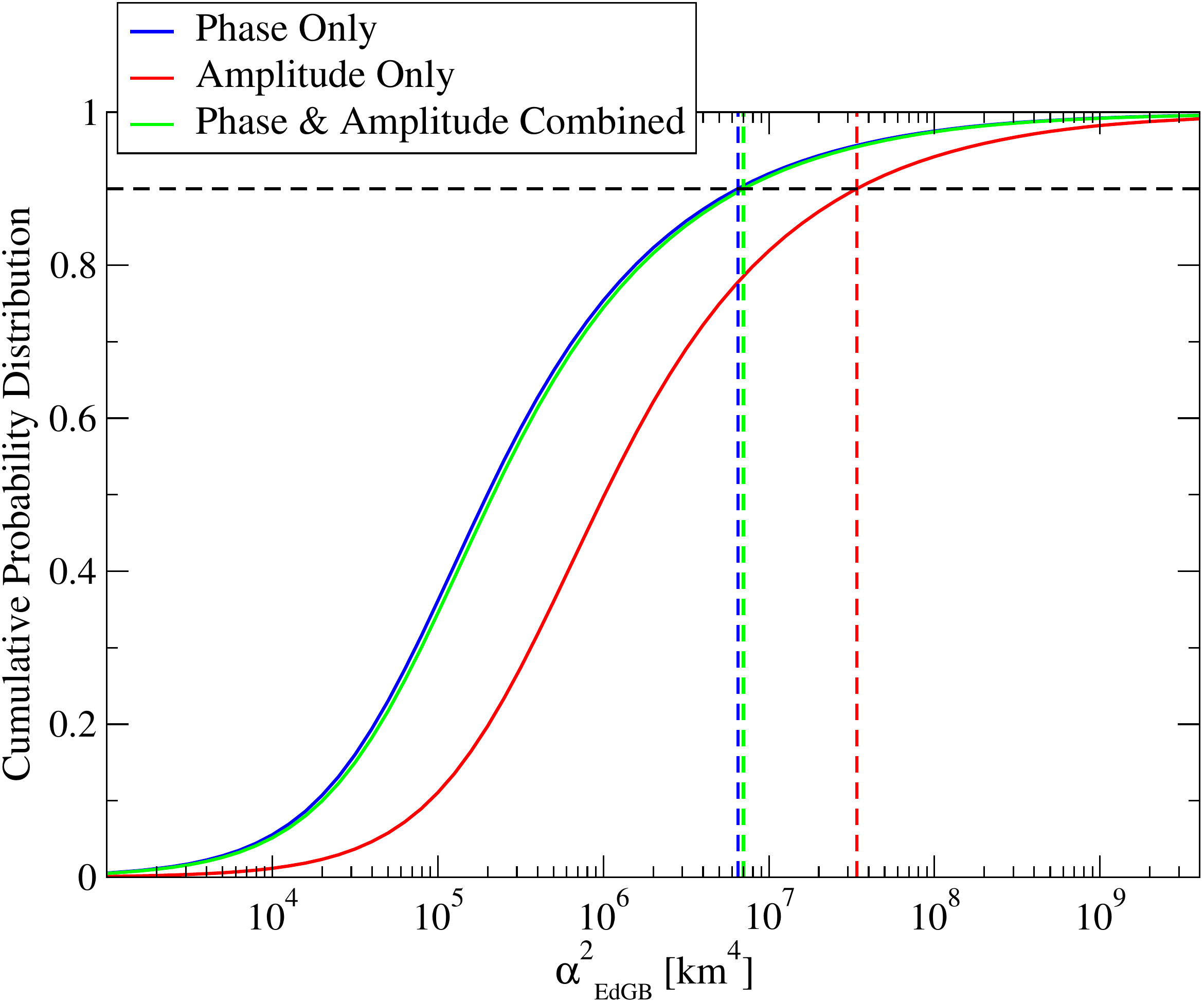}
\caption{Cumulative probability distributions of $\bar{\alpha}^2_{\EDGB}$ obtained from GW150914 for the same three cases as in Fig.~\ref{fig:histogram-edgb}. Each vertical dashed line shows the corresponding 90\% CL upper bound of a solid line of the same color.}
\label{fig:pdf-edgb}
\end{figure}

Figure~\ref{fig:pdf-edgb} presents cumulative probability distributions of $\bar{\alpha}^2_{\EDGB}$\footnote{We show the distribution of $\bar{\alpha}^2_{\EDGB}$ instead of $\sqrt{\bar{\alpha}_{\EDGB}}$ as it is the former that directly enters in the waveform.} for GW150914 for three different cases with vertical lines representing the 90\% CL of the corresponding distribution. We found the 90\% CL constraints on  $\sqrt{\bar{\alpha}_{\EDGB}}$ from each of the phase and amplitude corrections as 50.5 km and 76.3 km respectively. Notice that these bounds have the same order of magnitude. On the other hand, combining the amplitude and phase corrections leads to an upper bound of 51.5 km, which is \emph{weaker} than the  phase-only constraint by 2\% (to be discussed more in Sec.~\ref{conclusion}). 
Though above constraints may not be reliable as the 90\% CL bounds on $\zeta_\EDGB$ do not satisfy the small coupling approximation, which is shown in Table~\ref{table:ppE}.

We now look at bounds on GW151226. We found that this event yields 4.32 km and 10.5 km respectively from the phase and amplitude corrections, while combining the two only changes the result from the phase-only case by 0.01\%. These bounds are consistent with those in a recent paper~\cite{Nair:2019iur} that utilized the LVC posterior samples including the non-GR phase corrections at $-1$PN order while Ref.~\cite{Yamada:2019zrb} found even stronger bounds by combining multiple GW events. These GW bounds are comparable to the one obtained from low-mass X-ray binaries~\cite{Yagi:2012gp}. 

Although GW150914 leads to weaker constraints on EdGB gravity compared to GW151226, the effect of amplitude correction is more manifest for the former event. This is because GW150914 has a smaller number of GW cycles, and thus the amplitude contribution becomes relatively higher than GW151226.


\subsection{Scalar-Tensor Theories}
Scalar-Tensor theories of gravity emerge from the dimensional reduction of higher-dimensional theories such as the Kaluza-Klein theory~\cite{Fujii:2003pa,Overduin:1998pn} and string theories~\cite{polchinski1,polchinski2}. In addition to the spacetime curvature, scalar fields mediate an additional force which is introduced through non-minimal couplings of scalar fields and gravity~\cite{Berti:2015itd,Chiba:1997ms,PhysRevD.6.2077}. Such theories can explain the accelerated expansion of the universe~\cite{Brax:2004qh,PhysRevD.73.083510,PhysRevD.62.123510,PhysRevD.66.023525,Schimd:2004nq}, inflation~\cite{Burd:1991ns,Barrow:1990nv,Clifton:2011jh}, primordial nucleosynthesis~\cite{Coc:2006rt,Damour:1998ae,Larena:2005tu,Torres:1995je}, and the structure formation~\cite{Brax:2005ew}. 

Certain scalar-tensor theories predict scalarization of neutron stars~\cite{PhysRevLett.70.2220,Barausse:2012da}, which can also happen to BHs if the scalar field evolves with time cosmologically~\cite{Jacobson:1999vr,Horbatsch:2011ye}. Compact binaries formed by such objects emit dipole radiation which modifies the GW phase with~\cite{Freire:2012mg,Wex:2014nva,Tahura:2018zuq}
\be\label{eq:betaST}
\beta_{\ST}=-\frac{5}{7168}\eta ^{2/5}(\alpha_1-\alpha_2)^2
\ee
with $b=-7$, and the GW amplitude as~\cite{Tahura:2018zuq}
\be\label{eq:alphaST}
\alpha_{\ST}=-\frac{5}{192}\eta ^{2/5}(\alpha_1-\alpha_2)^2
\ee
with $a=-2$. Here $\alpha_A$ represents the scalar charge of the $A$th binary component and depends on specific theories and the type of compact objects. If we consider a binary consisting of BHs in a theory where the scalar field $\phi$ obeys a massless Klein-Gordon equation, $\alpha_A$ is given by~\cite{Horbatsch:2011ye}
\be\label{eq:alpha_A}
\alpha_A = 2 \, m_A \, \dot \phi\, [1+(1-\chi_A^2)^{1/2}]\,,
\ee
where $\dot{\phi}$ is the rate of change of $\phi$ with time.

One can use Eqs.~\eqref{eq:betaST}--\eqref{eq:alpha_A} and the numerical analysis described in Sec.~\ref{data} to find constraints on $\dot{\phi}$ as long as the small coupling approximation $m_A\dot{\phi}<1$ is satisfied. In this regard, only 11.7\% (13.5\%) of the samples of GW150914 satisfies such approximation with 90\% confidence level for the phase (combined) correction, while all of the samples fail to do so for the amplitude correction. Hence GW150914 cannot place any meaningful bound on scalar-tensor theories considered here. On the other hand, 90.4\% of the samples from GW151226 meets the small coupling criterion for the phase-only and combined analyses, while the fraction is only 25\% for the amplitude correction. Thus, we derive reliable constraints from GW151226 with the phase-only and combined analyses, with both leading to $\dot{\phi}<1.1\times10^4$/sec\footnote{A previous analysis with the sky-averaged waveform in Ref.~\cite{Yunes:2016jcc} could not place a reliable bound on scalar-tensor theories. Since the posterior distributions of the GW events were not available then, one could not determine how well those events satisfied the small coupling approximation from a simple Fisher analysis.}. This constraint is 10 orders of magnitude weaker than the current most stringent bound obtained from the orbital decay rate of quasar OJ287~\cite{Horbatsch:2011ye}, though this is the first bound obtained in the strong/dynamical regime.

\subsection{Varying- $G$ Theories}\label{vaying-G}
Many metric theories of gravity that violate the strong equivalence principle~\cite{DiCasola:2013iia,Will:2014kxa,0264-9381-7-10-007} predict time variation in the gravitational coupling parameter $G$~\cite{uzan:2010pm}. Scalar-tensor theories are examples where $G$ varies as a function of the asymptotic scalar field~\cite{Will2006}, which may vary over time. Any such time-dependence of $G$ leads to a variation in the effective masses of compact bodies, which in turn makes them experience anomalous cosmic acceleration~\cite{PhysRevLett.65.953}. Such phenomena alter the gravitational waveform through the modifications of the binary orbital evolution and the energy balance law~\cite{Tahura:2018zuq}.

We now show the PPE modifications due to a time variation in the gravitational constant. In fact, the amount of gravitational coupling that appears in different sectors of a  gravitational theory may not be unique. Einstein-\AE ther theory~\cite{Yagi:2013ava} and Brans-Dicke theory with a cosmologically evolving scalar field~\cite{Will2006} are examples of such theories in which various gravitational constants exist. Reference~\cite{Tahura:2018zuq} studied a generic case with two distinct gravitational constants in Kepler's law (conservative sector) and GW luminosity (dissipative sector). Here we place constraints on the special case where these two constants coincide with each other. Let the masses and the Newton's constant vary according to the following equations:
 \begin{eqnarray}\label{eq:gdot-1}
 m_A(t)\approx m_{A,0}+\dot{m}_{A,0}(t-t_0)\,, \\\label{eq:gdot-2}
   \label{eq:3.7a4}  G(t)\approx  G_{0}+\dot{G}_{0}(t-t_0)\,.
 \end{eqnarray}
 Here, a subscript $0$ denotes that the quantity is measured at the time $t=t_0$, and an overhead dot means a derivative with respect to time. Eqs.~\eqref{eq:gdot-1} and~\eqref{eq:gdot-2} modify the GW phase and amplitude as~\cite{Tahura:2018zuq}
 \begin{eqnarray}\label{eq:beta_gdot}
 \beta_{\dot{G}}&=&-\frac{25}{851968}\dot{G}_{0}\,\eta_0^{3/5}\left[(11+3s_1+3s_2)m_0 \right. \nonumber \\
 && \left. -41 (s_{1}m_{1}+s_{2}m_{2})\right]\,
  \end{eqnarray} 
with $b=-13$, and 
\begin{eqnarray}\label{eq:alpha_gdot}
\label{eq:3.7d2}
 \alpha_{\dot{G}}&=&\frac{5}{512}\eta_0^{3/5}\dot{G}_{0}\left[-(7-s_1-s_2)m_0 \right. \nonumber \\
 && \left.+13 (s_{1}m_{1}+s_{2}m_{2})\right]\, 
 \end{eqnarray}
with $a=-8$ respectively. Here $s_A$ is the sensitivity of the $A$th binary component defined as
\be
s_A=-\frac{G}{m_A}\frac{\partial m_A}{\partial  G}\bigg|_{t_0}\,.
\ee
 
Employing Eqs.~\eqref{eq:beta_gdot} and~\eqref{eq:alpha_gdot}, GW150914 (GW151226) imposes constraints on $|\dot{G_0}/G_0|$ from the phase-only and amplitude-only analyses as $7.30\times10^6\, \mathrm{yr}^{-1}$ ($2.24\times10^4\, \mathrm{yr}^{-1}$) and $1.37\times10^8\, \mathrm{yr}^{-1}$ ($3.82\times10^5\, \mathrm{yr}^{-1}$) respectively, with the combined analyses yielding slight improvements over the phase-only results.
Unlike the EdGB and scalar-tensor cases, the amplitude-only analyses yield much worse bound than that from the phase-only cases even with GW150914. Notice also that for varying-$G$ theories, the combined bound is slightly \emph{stronger} than the phase-only bound (to be discussed more in the subsequent section).
These bounds are much less stringent compared to the other contemporary constraints~\cite{Will2006}. 
However, future space-borne detectors such as LISA~\cite{Seoane:2013qna,Audley:2017drz} will be able to obtain constraints up to 13 orders of magnitude stronger compared to  the aLIGO ones~\cite{Yunes:2009bv,Chamberlain:2017fjl}.

\section{Conclusion and Discussion}\label{conclusion}
In this analysis, we have derived constraints on scalar-tensor, EdGB and varying-$G$ theories from GW150914 and GW151226. To do so, we performed Fisher analyses with Monte-Carlo simulations using the posterior samples constructed by LVC.
In particular, we derived reliable constraints on the time-evolution of the scalar field in scalar-tensor theories from GW observations for the first time. 

We explored how amplitude corrections contribute to the constraints on such theories. We derived three sets of bounds on each theory: phase-only, amplitude-only, and from both phase and amplitude combined. We found that for binaries with large masses such as GW150914, where we have less number of cycles, the bounds from the amplitude and phase can be comparable to each other. On the other hand, combined analyses yield constraints that differ from the phase-only case at most by 3.6\% for the theories under consideration. Hence, at least in theories where the leading corrections enter at negative PN orders, the phase-only analyses as done in previous literature~\cite{Yunes:2016jcc,Chamberlain:2017fjl,Nair:2019iur,Yamada:2019zrb,Carson:2019fxr} can produce sufficiently accurate constraints.
\begin{figure}[htb]
\includegraphics[width=8.5cm]{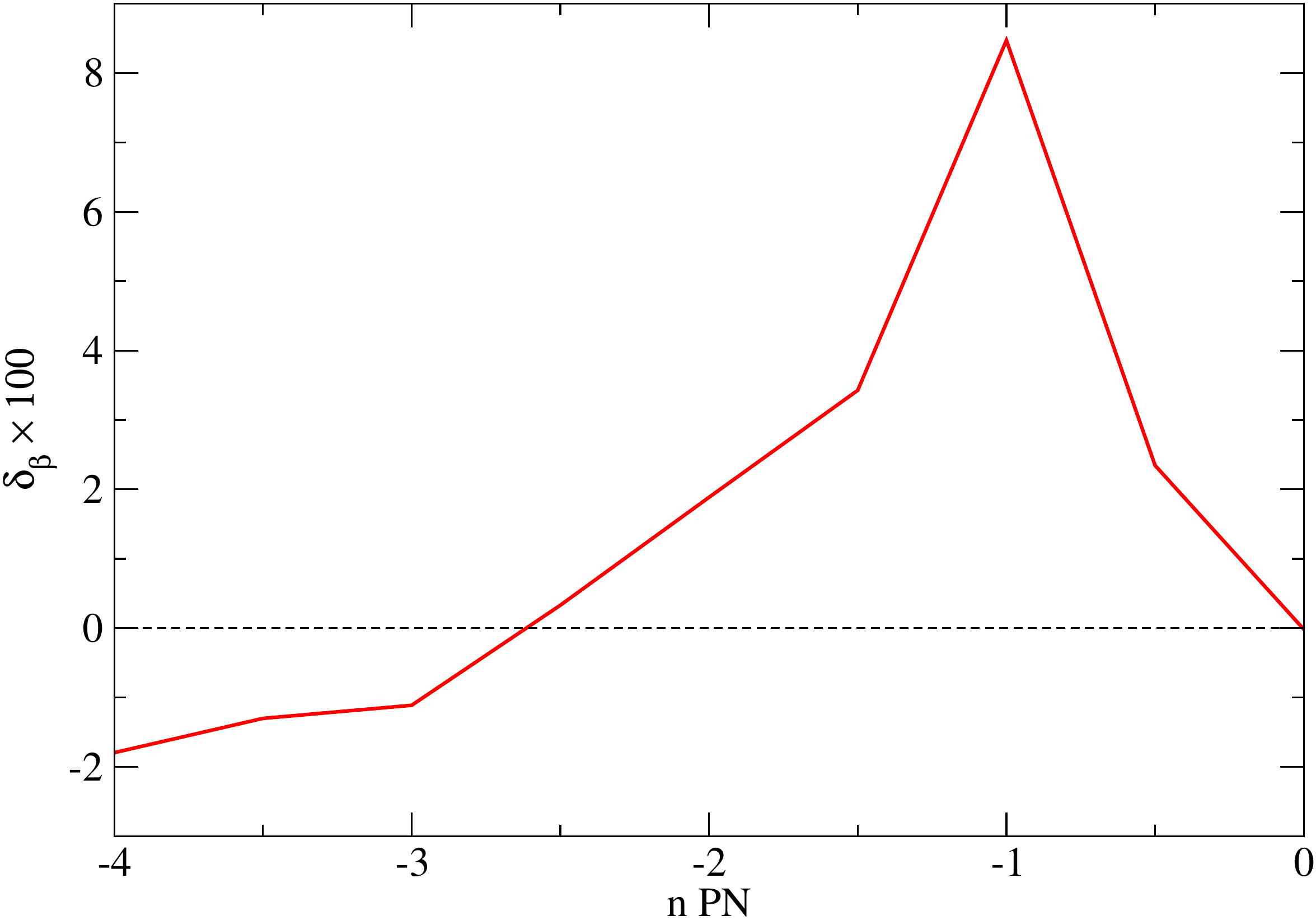}
\caption{Comparison of combined and phase-only analyses at different PN orders from GW150914 with a sky-averaged phenomB waveform. We show $\delta_{\beta}=\lb\beta_{\text{comb}}-\beta_{\text{phase}}\rb/\beta_{\text{phase}}$, where $\beta_{\text{phase}}$ and $\beta_{\text{comb}}$ are bounds on $\beta_{\PPE}$ from phase-only and combined analyses respectively. When $\delta_\beta$ is positive (negative), the combined analyses yield weaker (stronger) bounds than the phase-only ones.}
\label{fig:combinedvsphase}
\end{figure}

Depending on the prior information and the PN order of the non-GR correction, a combined analysis can yield stronger or weaker constraint compared to a phase-only one. With the priors mentioned in Sec.~\ref{data}, the fractional difference between $\beta_{\PPE}$ for the two cases is presented in Fig.~\ref{fig:combinedvsphase}. From $-4$ PN to $-2.5$ PN correction, the combined analyses give rise to slight improvements over the phase-only constraints, while for other cases, the former is weaker with a maximum deterioration of 8.5\% at $-1$ PN order. Nonetheless, it would be safer to include both phase and amplitude corrections in the analysis as a lack of the former in the waveform may cause systematic errors on GR parameters such as luminosity distance if non-GR corrections exist in nature.

{In this paper, we considered only the leading PN corrections in the inspiral part of the waveform, but how important are higher-PN corrections and modifications in the merger-ringdown portion? Reference~\cite{Yunes:2016jcc} partially addressed this question by taking Brans-Dicke theory as an example whose leading correction enters at $-1$PN order, similar to EdGB gravity and scalar-tensor theories considered here. Appendix~B of ~\cite{Yunes:2016jcc} shows that including higher-PN corrections only affects the bound from the leading PN correction by 10\% at most for GW150914. Moreover, for EdGB gravity, including the correction to the black hole ringdown frequency and damping time only affects the bound from the leading PN corrections in the inspiral by 4.5\% for GW150914~\cite{carsonyagi:2019}. Thus, it is likely that the bounds presented here are valid as order-of-magnitude estimates.}

A possible avenue for future work includes repeating the calculation presented here but with a Bayesian analysis using a more accurate waveform such as PhenomD, PhenomPv2 or effective-one-body ones. In particular, it would be interesting to investigate whether the amplitude correction contribution entering at positive PN orders is negligible like the negative PN cases reported here. {It is also interesting to repeat the analysis here to all the other events in GWTC-1~\cite{LIGOScientific:2018mvr} and study how much the bounds on each theory improve by combining these events.} Another possibility is to take into account non-tensorial polarization modes following e.g.~\cite{Chatziioannou:2012rf}. For future detectors with improved sensitivities, the ability to measure amplitude corrections may be dominated by calibration errors\footnote{The calibration error on the amplitude for the O2 run was 3.8\%~\cite{LIGOScientific:2018mvr}.}.

\acknowledgments

We thank Carl-Johan Haster for providing valuable comments on the LIGO posterior samples. 
K.Y. and Z.C. acknowledge support from NSF Award PHY-1806776. 
K.Y. would like to also acknowledge support by the COST Action GWverse CA16104 and JSPS KAKENHI Grants No. JP17H06358.

 \appendix 
\section{Comparison of Bounds on PPE parameters with PhenomB and PhenomD Waveforms}\label{Appendix}

Even though the PhenomD waveform produces more accurate results, we utilized the PhenomB one in this paper throughout because the latter is simpler and saves computational time when performing Monte Carlo simulations. In this appendix, we compare constraints on the PPE parameter $\alpha_\PPE$ from both waveforms to justify our method. 
 
Let us discuss the distinct features of the two waveforms first. Both PhenomB and PhenomD waveforms are spin-aligned (non-precessing) frequency-domain phenomenological models of gravitational waveforms~\cite{Ajith:2009bn,Khan:2015jqa}. The PhenomB waveform is calibrated for mass ratios up to $m_1/m_2 = 4$ and spin components of $\chi_i\in[-0.85,0.85]$ are unified into a single effective spin. On the other hand, the PhenomD waveform covers a larger region of the parameter space with mass ratios upto 18 and spins of $\chi_i\in[-0.95,0.95]$, with both spins introduced independently. The waveform contains a much higher order in PN terms in the inspiral than the PhenomB waveform and further introduces an intermediate phase connecting the inspiral and merger-ringdown portions,
which make such waveforms more reliable than the PhenomB ones.

\begin{figure}[htb]
\includegraphics[width=8.5cm]{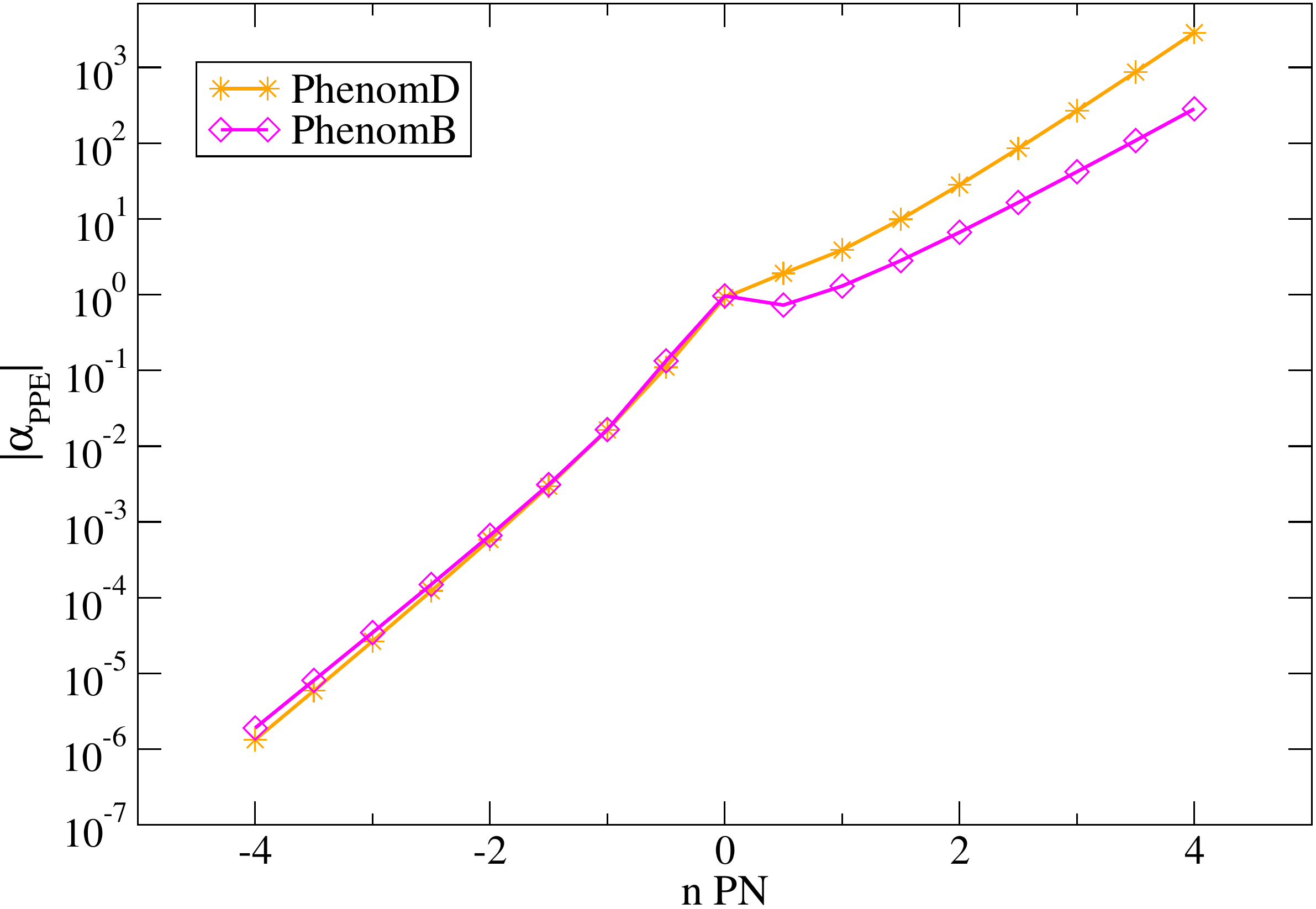}
\caption{Comparison of 90\% confidence constraints on $\alpha_{\PPE}$ from GW1501914 with the PhenomB and PhenomD waveforms for generation effects. }
\label{fig:phenomBvsD}
\end{figure}

\begin{figure}[htb]
\includegraphics[width=8.5cm]{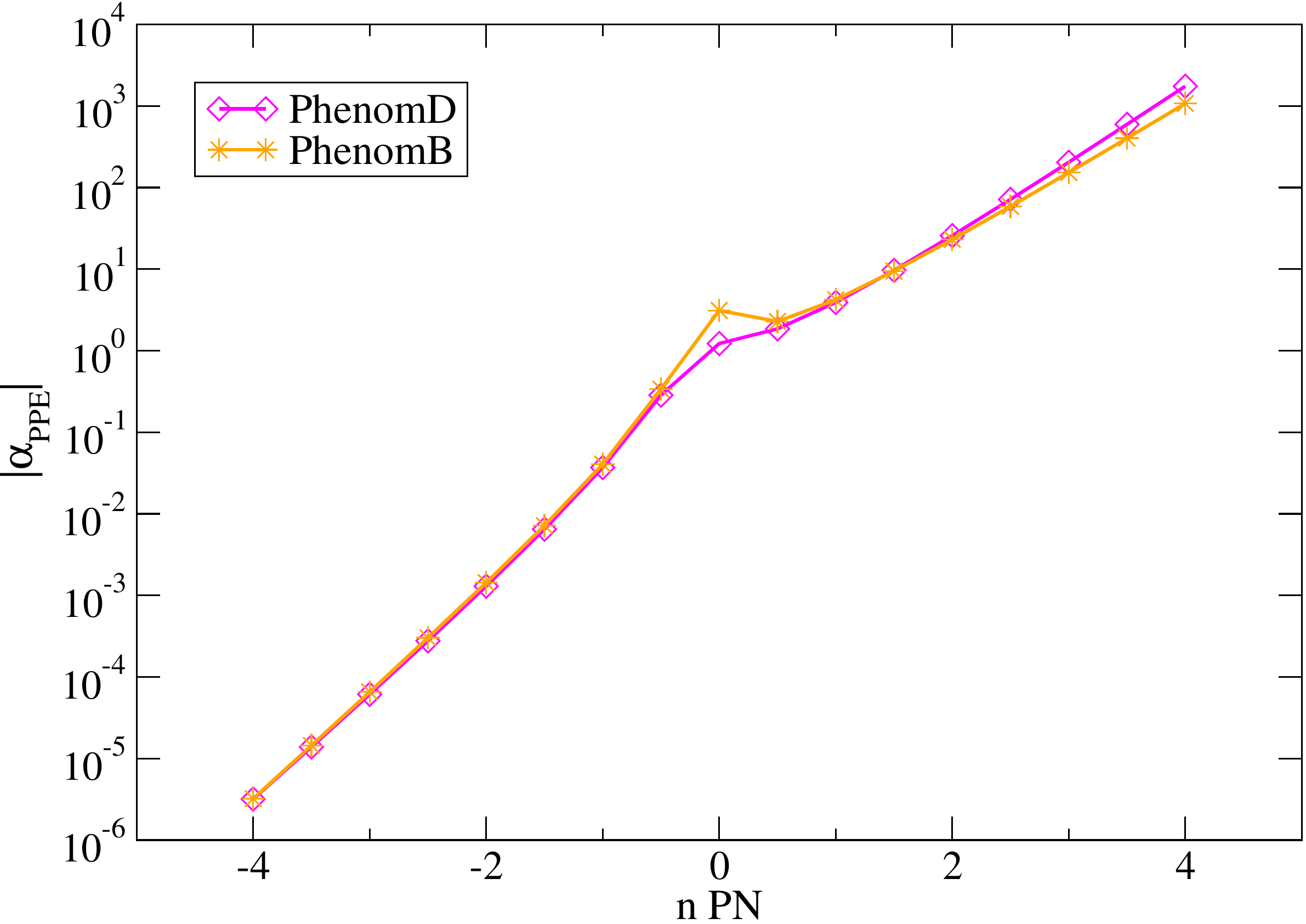}
\caption{Similar to Fig.~\ref{fig:phenomBvsD} but with inspiral signals only. The Fisher analyses are truncated at 104Hz which is corresponding to the transition frequency between the inspiral and merger portions of the PhenomB waveform, and we use the inspiral portion of the PhenomD waveform all the way up to this cutoff frequency.}
\label{fig:phenomBvsD-insp}
\end{figure}

We now estimate the constraints on the PPE amplitude modification from the two waveforms. Since modifications to propagation mechanisms used for massive gravity in Sec.~\ref{sec:massive} do not give rise to amplitude corrections, we here focus on modifications to generation mechanisms. We performed Fisher analyses with sky-averaged PhenomB and PhenomD waveforms and derived upper bounds on $\alpha_{\PPE}$ at different PN orders. As shown in Fig.~\ref{fig:phenomBvsD}, the results from the two waveforms agree very well at negative PN corrections but deviate from each other at the positive ones. 
On the other hand, truncating the Fisher analyses at the end of the inspiral phase show significant agreement between the two waveforms at positive PN orders (Fig.~\ref{fig:phenomBvsD-insp}), suggesting that the deviation in Fig.~\ref{fig:phenomBvsD} originates mainly from the intermediate/merger-ringdown portion.

The example theories considered in this paper acquire leading non-GR corrections either from propagation effects or from the generation effects with the latter entering in negative PN orders. A comparison between the PhenomB and PhenomD results for constraining $\beta_{\PPE}$ presented in Ref.~\cite{Yunes:2016jcc} reveals consistency on constraining modifications to propagation mechanisms at both positive and negative PN orders, while the two waveforms show agreement only at negative PN orders for constraining modifications to generation mechanisms. Together with the results on amplitude corrections discussed above confirms that the results of this paper should not change significantly if one utilizes the PhenomD waveform instead. On the other hand, for constraining theories like dynamical Chern-Simons or noncommutative gravity where the leading correction enters at a positive PN order, the PhenomB waveform is not expected to produce reliable results.

\bibliography{ppE-Numerical}

\begin{thebibliography}{122}%
\makeatletter
\providecommand \@ifxundefined [1]{%
 \@ifx{#1\undefined}
}%
\providecommand \@ifnum [1]{%
 \ifnum #1\expandafter \@firstoftwo
 \else \expandafter \@secondoftwo
 \fi
}%
\providecommand \@ifx [1]{%
 \ifx #1\expandafter \@firstoftwo
 \else \expandafter \@secondoftwo
 \fi
}%
\providecommand \natexlab [1]{#1}%
\providecommand \enquote  [1]{``#1''}%
\providecommand \bibnamefont  [1]{#1}%
\providecommand \bibfnamefont [1]{#1}%
\providecommand \citenamefont [1]{#1}%
\providecommand \href@noop [0]{\@secondoftwo}%
\providecommand \href [0]{\begingroup \@sanitize@url \@href}%
\providecommand \@href[1]{\@@startlink{#1}\@@href}%
\providecommand \@@href[1]{\endgroup#1\@@endlink}%
\providecommand \@sanitize@url [0]{\catcode `\\12\catcode `\$12\catcode
  `\&12\catcode `\#12\catcode `\^12\catcode `\_12\catcode `\%12\relax}%
\providecommand \@@startlink[1]{}%
\providecommand \@@endlink[0]{}%
\providecommand \url  [0]{\begingroup\@sanitize@url \@url }%
\providecommand \@url [1]{\endgroup\@href {#1}{\urlprefix }}%
\providecommand \urlprefix  [0]{URL }%
\providecommand \Eprint [0]{\href }%
\providecommand \doibase [0]{http://dx.doi.org/}%
\providecommand \selectlanguage [0]{\@gobble}%
\providecommand \bibinfo  [0]{\@secondoftwo}%
\providecommand \bibfield  [0]{\@secondoftwo}%
\providecommand \translation [1]{[#1]}%
\providecommand \BibitemOpen [0]{}%
\providecommand \bibitemStop [0]{}%
\providecommand \bibitemNoStop [0]{.\EOS\space}%
\providecommand \EOS [0]{\spacefactor3000\relax}%
\providecommand \BibitemShut  [1]{\csname bibitem#1\endcsname}%
\let\auto@bib@innerbib\@empty
\bibitem [{\citenamefont {Will}(2014)}]{Will:2014kxa}%
  \BibitemOpen
  \bibfield  {author} {\bibinfo {author} {\bibfnamefont {C.~M.}\ \bibnamefont
  {Will}},\ }\href {\doibase 10.12942/lrr-2014-4} {\bibfield  {journal}
  {\bibinfo  {journal} {Living Rev. Rel.}\ }\textbf {\bibinfo {volume} {17}},\
  \bibinfo {pages} {4} (\bibinfo {year} {2014})},\ \Eprint
  {http://arxiv.org/abs/1403.7377} {arXiv:1403.7377 [gr-qc]} \BibitemShut
  {NoStop}%
\bibitem [{\citenamefont {Bosma}(1981{\natexlab{a}})}]{article}%
  \BibitemOpen
  \bibfield  {author} {\bibinfo {author} {\bibfnamefont {A.}~\bibnamefont
  {Bosma}},\ }\bibfield  {booktitle} {\emph {\bibinfo {booktitle} {The
  Astronomical Journal}},\ }\href@noop {} {\ \textbf {\bibinfo {volume} {86}},\
  \bibinfo {pages} {1791} (\bibinfo {year} {1981}{\natexlab{a}})}\BibitemShut
  {NoStop}%
\bibitem [{\citenamefont {Bosma}(1981{\natexlab{b}})}]{Bosma:1981zz}%
  \BibitemOpen
  \bibfield  {author} {\bibinfo {author} {\bibfnamefont {A.}~\bibnamefont
  {Bosma}},\ }\href {\doibase 10.1086/113063} {\bibfield  {journal} {\bibinfo
  {journal} {Astron. J.}\ }\textbf {\bibinfo {volume} {86}},\ \bibinfo {pages}
  {1825} (\bibinfo {year} {1981}{\natexlab{b}})}\BibitemShut {NoStop}%
\bibitem [{\citenamefont {Begeman}\ \emph {et~al.}(1991)\citenamefont
  {Begeman}, \citenamefont {Broeils},\ and\ \citenamefont
  {Sanders}}]{Begeman:1991iy}%
  \BibitemOpen
  \bibfield  {author} {\bibinfo {author} {\bibfnamefont {K.~G.}\ \bibnamefont
  {Begeman}}, \bibinfo {author} {\bibfnamefont {A.~H.}\ \bibnamefont
  {Broeils}}, \ and\ \bibinfo {author} {\bibfnamefont {R.~H.}\ \bibnamefont
  {Sanders}},\ }\href@noop {} {\bibfield  {journal} {\bibinfo  {journal} {Mon.
  Not. Roy. Astron. Soc.}\ }\textbf {\bibinfo {volume} {249}},\ \bibinfo
  {pages} {523} (\bibinfo {year} {1991})}\BibitemShut {NoStop}%
\bibitem [{\citenamefont {Rubin}\ and\ \citenamefont
  {Ford}(1970)}]{Rubin:1970zza}%
  \BibitemOpen
  \bibfield  {author} {\bibinfo {author} {\bibfnamefont {V.~C.}\ \bibnamefont
  {Rubin}}\ and\ \bibinfo {author} {\bibfnamefont {W.~K.}\ \bibnamefont {Ford},
  \bibfnamefont {Jr.}},\ }\href {\doibase 10.1086/150317} {\bibfield  {journal}
  {\bibinfo  {journal} {Astrophys. J.}\ }\textbf {\bibinfo {volume} {159}},\
  \bibinfo {pages} {379} (\bibinfo {year} {1970})}\BibitemShut {NoStop}%
\bibitem [{\citenamefont {Rubin}\ \emph {et~al.}(1980)\citenamefont {Rubin},
  \citenamefont {Thonnard},\ and\ \citenamefont {Ford}}]{Rubin:1980zd}%
  \BibitemOpen
  \bibfield  {author} {\bibinfo {author} {\bibfnamefont {V.~C.}\ \bibnamefont
  {Rubin}}, \bibinfo {author} {\bibfnamefont {N.}~\bibnamefont {Thonnard}}, \
  and\ \bibinfo {author} {\bibfnamefont {W.~K.}\ \bibnamefont {Ford},
  \bibfnamefont {Jr.}},\ }\href {\doibase 10.1086/158003} {\bibfield  {journal}
  {\bibinfo  {journal} {Astrophys. J.}\ }\textbf {\bibinfo {volume} {238}},\
  \bibinfo {pages} {471} (\bibinfo {year} {1980})}\BibitemShut {NoStop}%
\bibitem [{\citenamefont {{Ostriker}}\ and\ \citenamefont
  {{Peebles}}(1973)}]{1973ApJ...186..467O}%
  \BibitemOpen
  \bibfield  {author} {\bibinfo {author} {\bibfnamefont {J.~P.}\ \bibnamefont
  {{Ostriker}}}\ and\ \bibinfo {author} {\bibfnamefont {P.~J.~E.}\ \bibnamefont
  {{Peebles}}},\ }\href {\doibase 10.1086/152513} {\bibfield  {journal}
  {\bibinfo  {journal} {\apj}\ }\textbf {\bibinfo {volume} {186}},\ \bibinfo
  {pages} {467} (\bibinfo {year} {1973})}\BibitemShut {NoStop}%
\bibitem [{\citenamefont {Ostriker}(1993)}]{Ostriker:1993fr}%
  \BibitemOpen
  \bibfield  {author} {\bibinfo {author} {\bibfnamefont {J.~P.}\ \bibnamefont
  {Ostriker}},\ }\href {\doibase 10.1146/annurev.aa.31.090193.003353}
  {\bibfield  {journal} {\bibinfo  {journal} {Ann. Rev. Astron. Astrophys.}\
  }\textbf {\bibinfo {volume} {31}},\ \bibinfo {pages} {689} (\bibinfo {year}
  {1993})}\BibitemShut {NoStop}%
\bibitem [{\citenamefont {Abbott}(1988)}]{Abbott:1988nx}%
  \BibitemOpen
  \bibfield  {author} {\bibinfo {author} {\bibfnamefont {L.}~\bibnamefont
  {Abbott}},\ }\href {\doibase 10.1038/scientificamerican0588-106} {\bibfield
  {journal} {\bibinfo  {journal} {Sci. Am.}\ }\textbf {\bibinfo {volume}
  {258}},\ \bibinfo {pages} {106} (\bibinfo {year} {1988})},\ \bibinfo {note}
  {[Spektrum Wiss.7,92(1988)]}\BibitemShut {NoStop}%
\bibitem [{\citenamefont {Copeland}\ \emph {et~al.}(2006)\citenamefont
  {Copeland}, \citenamefont {Sami},\ and\ \citenamefont
  {Tsujikawa}}]{Copeland:2006wr}%
  \BibitemOpen
  \bibfield  {author} {\bibinfo {author} {\bibfnamefont {E.~J.}\ \bibnamefont
  {Copeland}}, \bibinfo {author} {\bibfnamefont {M.}~\bibnamefont {Sami}}, \
  and\ \bibinfo {author} {\bibfnamefont {S.}~\bibnamefont {Tsujikawa}},\ }\href
  {\doibase 10.1142/S021827180600942X} {\bibfield  {journal} {\bibinfo
  {journal} {Int. J. Mod. Phys.}\ }\textbf {\bibinfo {volume} {D15}},\ \bibinfo
  {pages} {1753} (\bibinfo {year} {2006})},\ \Eprint
  {http://arxiv.org/abs/hep-th/0603057} {arXiv:hep-th/0603057 [hep-th]}
  \BibitemShut {NoStop}%
\bibitem [{\citenamefont {Perlmutter}\ \emph {et~al.}(1999)\citenamefont
  {Perlmutter} \emph {et~al.}}]{Perlmutter:1998np}%
  \BibitemOpen
  \bibfield  {author} {\bibinfo {author} {\bibfnamefont {S.}~\bibnamefont
  {Perlmutter}} \emph {et~al.} (\bibinfo {collaboration} {Supernova Cosmology
  Project}),\ }\href {\doibase 10.1086/307221} {\bibfield  {journal} {\bibinfo
  {journal} {Astrophys. J.}\ }\textbf {\bibinfo {volume} {517}},\ \bibinfo
  {pages} {565} (\bibinfo {year} {1999})},\ \Eprint
  {http://arxiv.org/abs/astro-ph/9812133} {arXiv:astro-ph/9812133 [astro-ph]}
  \BibitemShut {NoStop}%
\bibitem [{\citenamefont {Riess}\ \emph {et~al.}(1998)\citenamefont {Riess}
  \emph {et~al.}}]{Riess:1998cb}%
  \BibitemOpen
  \bibfield  {author} {\bibinfo {author} {\bibfnamefont {A.~G.}\ \bibnamefont
  {Riess}} \emph {et~al.} (\bibinfo {collaboration} {Supernova Search Team}),\
  }\href {\doibase 10.1086/300499} {\bibfield  {journal} {\bibinfo  {journal}
  {Astron. J.}\ }\textbf {\bibinfo {volume} {116}},\ \bibinfo {pages} {1009}
  (\bibinfo {year} {1998})},\ \Eprint {http://arxiv.org/abs/astro-ph/9805201}
  {arXiv:astro-ph/9805201 [astro-ph]} \BibitemShut {NoStop}%
\bibitem [{\citenamefont {Riess}\ \emph {et~al.}(2004)\citenamefont {Riess}
  \emph {et~al.}}]{Riess:2004nr}%
  \BibitemOpen
  \bibfield  {author} {\bibinfo {author} {\bibfnamefont {A.~G.}\ \bibnamefont
  {Riess}} \emph {et~al.} (\bibinfo {collaboration} {Supernova Search Team}),\
  }\href {\doibase 10.1086/383612} {\bibfield  {journal} {\bibinfo  {journal}
  {Astrophys. J.}\ }\textbf {\bibinfo {volume} {607}},\ \bibinfo {pages} {665}
  (\bibinfo {year} {2004})},\ \Eprint {http://arxiv.org/abs/astro-ph/0402512}
  {arXiv:astro-ph/0402512 [astro-ph]} \BibitemShut {NoStop}%
\bibitem [{\citenamefont {Weinberg}(1989)}]{RevModPhys.61.1}%
  \BibitemOpen
  \bibfield  {author} {\bibinfo {author} {\bibfnamefont {S.}~\bibnamefont
  {Weinberg}},\ }\href {\doibase 10.1103/RevModPhys.61.1} {\bibfield  {journal}
  {\bibinfo  {journal} {Rev. Mod. Phys.}\ }\textbf {\bibinfo {volume} {61}},\
  \bibinfo {pages} {1} (\bibinfo {year} {1989})}\BibitemShut {NoStop}%
\bibitem [{\citenamefont {van Albada}\ \emph {et~al.}(1985)\citenamefont {van
  Albada}, \citenamefont {Bahcall}, \citenamefont {Begeman},\ and\
  \citenamefont {Sancisi}}]{vanAlbada:1984js}%
  \BibitemOpen
  \bibfield  {author} {\bibinfo {author} {\bibfnamefont {T.~S.}\ \bibnamefont
  {van Albada}}, \bibinfo {author} {\bibfnamefont {J.~N.}\ \bibnamefont
  {Bahcall}}, \bibinfo {author} {\bibfnamefont {K.}~\bibnamefont {Begeman}}, \
  and\ \bibinfo {author} {\bibfnamefont {R.}~\bibnamefont {Sancisi}},\ }\href
  {\doibase 10.1086/163375} {\bibfield  {journal} {\bibinfo  {journal}
  {Astrophys. J.}\ }\textbf {\bibinfo {volume} {295}},\ \bibinfo {pages} {305}
  (\bibinfo {year} {1985})}\BibitemShut {NoStop}%
\bibitem [{\citenamefont {Weinberg}\ \emph {et~al.}(2013)\citenamefont
  {Weinberg}, \citenamefont {Mortonson}, \citenamefont {Eisenstein},
  \citenamefont {Hirata}, \citenamefont {Riess},\ and\ \citenamefont
  {Rozo}}]{WEINBERG201387}%
  \BibitemOpen
  \bibfield  {author} {\bibinfo {author} {\bibfnamefont {D.~H.}\ \bibnamefont
  {Weinberg}}, \bibinfo {author} {\bibfnamefont {M.~J.}\ \bibnamefont
  {Mortonson}}, \bibinfo {author} {\bibfnamefont {D.~J.}\ \bibnamefont
  {Eisenstein}}, \bibinfo {author} {\bibfnamefont {C.}~\bibnamefont {Hirata}},
  \bibinfo {author} {\bibfnamefont {A.~G.}\ \bibnamefont {Riess}}, \ and\
  \bibinfo {author} {\bibfnamefont {E.}~\bibnamefont {Rozo}},\ }\href {\doibase
  https://doi.org/10.1016/j.physrep.2013.05.001} {\bibfield  {journal}
  {\bibinfo  {journal} {Physics Reports}\ }\textbf {\bibinfo {volume} {530}},\
  \bibinfo {pages} {87 } (\bibinfo {year} {2013})},\ \bibinfo {note}
  {observational Probes of Cosmic Acceleration}\BibitemShut {NoStop}%
\bibitem [{\citenamefont {Adler}(2010)}]{Adler:2010wf}%
  \BibitemOpen
  \bibfield  {author} {\bibinfo {author} {\bibfnamefont {R.~J.}\ \bibnamefont
  {Adler}},\ }\href {\doibase 10.1119/1.3439650} {\bibfield  {journal}
  {\bibinfo  {journal} {Am. J. Phys.}\ }\textbf {\bibinfo {volume} {78}},\
  \bibinfo {pages} {925} (\bibinfo {year} {2010})},\ \Eprint
  {http://arxiv.org/abs/1001.1205} {arXiv:1001.1205 [gr-qc]} \BibitemShut
  {NoStop}%
\bibitem [{\citenamefont {Ng}(2003)}]{Ng:2003jk}%
  \BibitemOpen
  \bibfield  {author} {\bibinfo {author} {\bibfnamefont {Y.~J.}\ \bibnamefont
  {Ng}},\ }\href {\doibase 10.1142/S0217732303010934} {\bibfield  {journal}
  {\bibinfo  {journal} {Mod. Phys. Lett.}\ }\textbf {\bibinfo {volume} {A18}},\
  \bibinfo {pages} {1073} (\bibinfo {year} {2003})},\ \Eprint
  {http://arxiv.org/abs/gr-qc/0305019} {arXiv:gr-qc/0305019 [gr-qc]}
  \BibitemShut {NoStop}%
\bibitem [{\citenamefont {Abbott}\ \emph
  {et~al.}(2016{\natexlab{a}})\citenamefont {Abbott} \emph
  {et~al.}}]{TheLIGOScientific:2016src}%
  \BibitemOpen
  \bibfield  {author} {\bibinfo {author} {\bibfnamefont {B.~P.}\ \bibnamefont
  {Abbott}} \emph {et~al.} (\bibinfo {collaboration} {LIGO Scientific,
  Virgo}),\ }\href {\doibase 10.1103/PhysRevLett.116.221101,
  10.1103/PhysRevLett.121.129902} {\bibfield  {journal} {\bibinfo  {journal}
  {Phys. Rev. Lett.}\ }\textbf {\bibinfo {volume} {116}},\ \bibinfo {pages}
  {221101} (\bibinfo {year} {2016}{\natexlab{a}})},\ \bibinfo {note} {[Erratum:
  Phys. Rev. Lett.121,no.12,129902(2018)]},\ \Eprint
  {http://arxiv.org/abs/1602.03841} {arXiv:1602.03841 [gr-qc]} \BibitemShut
  {NoStop}%
\bibitem [{\citenamefont {Yunes}\ \emph {et~al.}(2016)\citenamefont {Yunes},
  \citenamefont {Yagi},\ and\ \citenamefont {Pretorius}}]{Yunes:2016jcc}%
  \BibitemOpen
  \bibfield  {author} {\bibinfo {author} {\bibfnamefont {N.}~\bibnamefont
  {Yunes}}, \bibinfo {author} {\bibfnamefont {K.}~\bibnamefont {Yagi}}, \ and\
  \bibinfo {author} {\bibfnamefont {F.}~\bibnamefont {Pretorius}},\ }\href
  {\doibase 10.1103/PhysRevD.94.084002} {\bibfield  {journal} {\bibinfo
  {journal} {Phys. Rev.}\ }\textbf {\bibinfo {volume} {D94}},\ \bibinfo {pages}
  {084002} (\bibinfo {year} {2016})},\ \Eprint
  {http://arxiv.org/abs/1603.08955} {arXiv:1603.08955 [gr-qc]} \BibitemShut
  {NoStop}%
\bibitem [{\citenamefont {Abbott}\ \emph {et~al.}(2019)\citenamefont {Abbott}
  \emph {et~al.}}]{LIGOScientific:2019fpa}%
  \BibitemOpen
  \bibfield  {author} {\bibinfo {author} {\bibfnamefont {B.~P.}\ \bibnamefont
  {Abbott}} \emph {et~al.} (\bibinfo {collaboration} {LIGO Scientific,
  Virgo}),\ }\href@noop {} {\  (\bibinfo {year} {2019})},\ \Eprint
  {http://arxiv.org/abs/1903.04467} {arXiv:1903.04467 [gr-qc]} \BibitemShut
  {NoStop}%
\bibitem [{\citenamefont {Abbott}\ \emph
  {et~al.}(2017{\natexlab{a}})\citenamefont {Abbott} \emph
  {et~al.}}]{Monitor:2017mdv}%
  \BibitemOpen
  \bibfield  {author} {\bibinfo {author} {\bibfnamefont {B.~P.}\ \bibnamefont
  {Abbott}} \emph {et~al.} (\bibinfo {collaboration} {Virgo, Fermi-GBM,
  INTEGRAL, LIGO Scientific}),\ }\href {\doibase 10.3847/2041-8213/aa920c}
  {\bibfield  {journal} {\bibinfo  {journal} {Astrophys. J.}\ }\textbf
  {\bibinfo {volume} {848}},\ \bibinfo {pages} {L13} (\bibinfo {year}
  {2017}{\natexlab{a}})},\ \Eprint {http://arxiv.org/abs/1710.05834}
  {arXiv:1710.05834 [astro-ph.HE]} \BibitemShut {NoStop}%
\bibitem [{\citenamefont {Abbott}\ \emph
  {et~al.}(2018{\natexlab{a}})\citenamefont {Abbott} \emph
  {et~al.}}]{Abbott:2018lct}%
  \BibitemOpen
  \bibfield  {author} {\bibinfo {author} {\bibfnamefont {B.~P.}\ \bibnamefont
  {Abbott}} \emph {et~al.} (\bibinfo {collaboration} {LIGO Scientific,
  Virgo}),\ }\href@noop {} {\  (\bibinfo {year} {2018}{\natexlab{a}})},\
  \Eprint {http://arxiv.org/abs/1811.00364} {arXiv:1811.00364 [gr-qc]}
  \BibitemShut {NoStop}%
\bibitem [{\citenamefont {Giddings}\ \emph {et~al.}(2019)\citenamefont
  {Giddings}, \citenamefont {Koren},\ and\ \citenamefont
  {Treviño}}]{Giddings:2019ujs}%
  \BibitemOpen
  \bibfield  {author} {\bibinfo {author} {\bibfnamefont {S.~B.}\ \bibnamefont
  {Giddings}}, \bibinfo {author} {\bibfnamefont {S.}~\bibnamefont {Koren}}, \
  and\ \bibinfo {author} {\bibfnamefont {G.}~\bibnamefont {Treviño}},\
  }\href@noop {} {\  (\bibinfo {year} {2019})},\ \Eprint
  {http://arxiv.org/abs/1904.04258} {arXiv:1904.04258 [gr-qc]} \BibitemShut
  {NoStop}%
\bibitem [{\citenamefont {Carballo-Rubio}\ \emph {et~al.}(2018)\citenamefont
  {Carballo-Rubio}, \citenamefont {Di~Filippo}, \citenamefont {Liberati},\ and\
  \citenamefont {Visser}}]{Carballo-Rubio:2018jzw}%
  \BibitemOpen
  \bibfield  {author} {\bibinfo {author} {\bibfnamefont {R.}~\bibnamefont
  {Carballo-Rubio}}, \bibinfo {author} {\bibfnamefont {F.}~\bibnamefont
  {Di~Filippo}}, \bibinfo {author} {\bibfnamefont {S.}~\bibnamefont
  {Liberati}}, \ and\ \bibinfo {author} {\bibfnamefont {M.}~\bibnamefont
  {Visser}},\ }\href {\doibase 10.1103/PhysRevD.98.124009} {\bibfield
  {journal} {\bibinfo  {journal} {Phys. Rev.}\ }\textbf {\bibinfo {volume}
  {D98}},\ \bibinfo {pages} {124009} (\bibinfo {year} {2018})},\ \Eprint
  {http://arxiv.org/abs/1809.08238} {arXiv:1809.08238 [gr-qc]} \BibitemShut
  {NoStop}%
\bibitem [{\citenamefont {Arun}\ \emph
  {et~al.}(2006{\natexlab{a}})\citenamefont {Arun}, \citenamefont {Iyer},
  \citenamefont {Qusailah},\ and\ \citenamefont {Sathyaprakash}}]{Arun:2006yw}%
  \BibitemOpen
  \bibfield  {author} {\bibinfo {author} {\bibfnamefont {K.~G.}\ \bibnamefont
  {Arun}}, \bibinfo {author} {\bibfnamefont {B.~R.}\ \bibnamefont {Iyer}},
  \bibinfo {author} {\bibfnamefont {M.~S.~S.}\ \bibnamefont {Qusailah}}, \ and\
  \bibinfo {author} {\bibfnamefont {B.~S.}\ \bibnamefont {Sathyaprakash}},\
  }\href {\doibase 10.1088/0264-9381/23/9/L01} {\bibfield  {journal} {\bibinfo
  {journal} {Class. Quant. Grav.}\ }\textbf {\bibinfo {volume} {23}},\ \bibinfo
  {pages} {L37} (\bibinfo {year} {2006}{\natexlab{a}})},\ \Eprint
  {http://arxiv.org/abs/gr-qc/0604018} {arXiv:gr-qc/0604018 [gr-qc]}
  \BibitemShut {NoStop}%
\bibitem [{\citenamefont {Arun}\ \emph
  {et~al.}(2006{\natexlab{b}})\citenamefont {Arun}, \citenamefont {Iyer},
  \citenamefont {Qusailah},\ and\ \citenamefont {Sathyaprakash}}]{Arun:2006hn}%
  \BibitemOpen
  \bibfield  {author} {\bibinfo {author} {\bibfnamefont {K.~G.}\ \bibnamefont
  {Arun}}, \bibinfo {author} {\bibfnamefont {B.~R.}\ \bibnamefont {Iyer}},
  \bibinfo {author} {\bibfnamefont {M.~S.~S.}\ \bibnamefont {Qusailah}}, \ and\
  \bibinfo {author} {\bibfnamefont {B.~S.}\ \bibnamefont {Sathyaprakash}},\
  }\href {\doibase 10.1103/PhysRevD.74.024006} {\bibfield  {journal} {\bibinfo
  {journal} {Phys. Rev.}\ }\textbf {\bibinfo {volume} {D74}},\ \bibinfo {pages}
  {024006} (\bibinfo {year} {2006}{\natexlab{b}})},\ \Eprint
  {http://arxiv.org/abs/gr-qc/0604067} {arXiv:gr-qc/0604067 [gr-qc]}
  \BibitemShut {NoStop}%
\bibitem [{\citenamefont {Mishra}\ \emph {et~al.}(2010)\citenamefont {Mishra},
  \citenamefont {Arun}, \citenamefont {Iyer},\ and\ \citenamefont
  {Sathyaprakash}}]{Mishra:2010tp}%
  \BibitemOpen
  \bibfield  {author} {\bibinfo {author} {\bibfnamefont {C.~K.}\ \bibnamefont
  {Mishra}}, \bibinfo {author} {\bibfnamefont {K.~G.}\ \bibnamefont {Arun}},
  \bibinfo {author} {\bibfnamefont {B.~R.}\ \bibnamefont {Iyer}}, \ and\
  \bibinfo {author} {\bibfnamefont {B.~S.}\ \bibnamefont {Sathyaprakash}},\
  }\href {\doibase 10.1103/PhysRevD.82.064010} {\bibfield  {journal} {\bibinfo
  {journal} {Phys. Rev.}\ }\textbf {\bibinfo {volume} {D82}},\ \bibinfo {pages}
  {064010} (\bibinfo {year} {2010})},\ \Eprint {http://arxiv.org/abs/1005.0304}
  {arXiv:1005.0304 [gr-qc]} \BibitemShut {NoStop}%
\bibitem [{\citenamefont {Yunes}\ and\ \citenamefont
  {Pretorius}(2009)}]{Yunes:2009ke}%
  \BibitemOpen
  \bibfield  {author} {\bibinfo {author} {\bibfnamefont {N.}~\bibnamefont
  {Yunes}}\ and\ \bibinfo {author} {\bibfnamefont {F.}~\bibnamefont
  {Pretorius}},\ }\href {\doibase 10.1103/PhysRevD.80.122003} {\bibfield
  {journal} {\bibinfo  {journal} {Phys. Rev.}\ }\textbf {\bibinfo {volume}
  {D80}},\ \bibinfo {pages} {122003} (\bibinfo {year} {2009})},\ \Eprint
  {http://arxiv.org/abs/0909.3328} {arXiv:0909.3328 [gr-qc]} \BibitemShut
  {NoStop}%
\bibitem [{\citenamefont {Chatziioannou}\ \emph {et~al.}(2012)\citenamefont
  {Chatziioannou}, \citenamefont {Yunes},\ and\ \citenamefont
  {Cornish}}]{Chatziioannou:2012rf}%
  \BibitemOpen
  \bibfield  {author} {\bibinfo {author} {\bibfnamefont {K.}~\bibnamefont
  {Chatziioannou}}, \bibinfo {author} {\bibfnamefont {N.}~\bibnamefont
  {Yunes}}, \ and\ \bibinfo {author} {\bibfnamefont {N.}~\bibnamefont
  {Cornish}},\ }\href {\doibase 10.1103/PhysRevD.86.022004,
  10.1103/PhysRevD.95.129901} {\bibfield  {journal} {\bibinfo  {journal} {Phys.
  Rev.}\ }\textbf {\bibinfo {volume} {D86}},\ \bibinfo {pages} {022004}
  (\bibinfo {year} {2012})},\ \bibinfo {note} {[Erratum: Phys.
  Rev.D95,no.12,129901(2017)]},\ \Eprint {http://arxiv.org/abs/1204.2585}
  {arXiv:1204.2585 [gr-qc]} \BibitemShut {NoStop}%
\bibitem [{\citenamefont {Agathos}\ \emph {et~al.}(2014)\citenamefont
  {Agathos}, \citenamefont {Del~Pozzo}, \citenamefont {Li}, \citenamefont {Van
  Den~Broeck}, \citenamefont {Veitch},\ and\ \citenamefont
  {Vitale}}]{Agathos:2013upa}%
  \BibitemOpen
  \bibfield  {author} {\bibinfo {author} {\bibfnamefont {M.}~\bibnamefont
  {Agathos}}, \bibinfo {author} {\bibfnamefont {W.}~\bibnamefont {Del~Pozzo}},
  \bibinfo {author} {\bibfnamefont {T.~G.~F.}\ \bibnamefont {Li}}, \bibinfo
  {author} {\bibfnamefont {C.}~\bibnamefont {Van Den~Broeck}}, \bibinfo
  {author} {\bibfnamefont {J.}~\bibnamefont {Veitch}}, \ and\ \bibinfo {author}
  {\bibfnamefont {S.}~\bibnamefont {Vitale}},\ }\href {\doibase
  10.1103/PhysRevD.89.082001} {\bibfield  {journal} {\bibinfo  {journal} {Phys.
  Rev.}\ }\textbf {\bibinfo {volume} {D89}},\ \bibinfo {pages} {082001}
  (\bibinfo {year} {2014})},\ \Eprint {http://arxiv.org/abs/1311.0420}
  {arXiv:1311.0420 [gr-qc]} \BibitemShut {NoStop}%
\bibitem [{\citenamefont {Meidam}\ \emph {et~al.}(2014)\citenamefont {Meidam},
  \citenamefont {Agathos}, \citenamefont {Van Den~Broeck}, \citenamefont
  {Veitch},\ and\ \citenamefont {Sathyaprakash}}]{Meidam:2014jpa}%
  \BibitemOpen
  \bibfield  {author} {\bibinfo {author} {\bibfnamefont {J.}~\bibnamefont
  {Meidam}}, \bibinfo {author} {\bibfnamefont {M.}~\bibnamefont {Agathos}},
  \bibinfo {author} {\bibfnamefont {C.}~\bibnamefont {Van Den~Broeck}},
  \bibinfo {author} {\bibfnamefont {J.}~\bibnamefont {Veitch}}, \ and\ \bibinfo
  {author} {\bibfnamefont {B.~S.}\ \bibnamefont {Sathyaprakash}},\ }\href
  {\doibase 10.1103/PhysRevD.90.064009} {\bibfield  {journal} {\bibinfo
  {journal} {Phys. Rev.}\ }\textbf {\bibinfo {volume} {D90}},\ \bibinfo {pages}
  {064009} (\bibinfo {year} {2014})},\ \Eprint {http://arxiv.org/abs/1406.3201}
  {arXiv:1406.3201 [gr-qc]} \BibitemShut {NoStop}%
\bibitem [{\citenamefont {Abbott}\ \emph
  {et~al.}(2016{\natexlab{b}})\citenamefont {Abbott} \emph
  {et~al.}}]{TheLIGOScientific:2016pea}%
  \BibitemOpen
  \bibfield  {author} {\bibinfo {author} {\bibfnamefont {B.~P.}\ \bibnamefont
  {Abbott}} \emph {et~al.} (\bibinfo {collaboration} {Virgo, LIGO
  Scientific}),\ }\href {\doibase 10.1103/PhysRevX.6.041015} {\bibfield
  {journal} {\bibinfo  {journal} {Phys. Rev.}\ }\textbf {\bibinfo {volume}
  {X6}},\ \bibinfo {pages} {041015} (\bibinfo {year} {2016}{\natexlab{b}})},\
  \Eprint {http://arxiv.org/abs/1606.04856} {arXiv:1606.04856 [gr-qc]}
  \BibitemShut {NoStop}%
\bibitem [{\citenamefont {Abbott}\ \emph
  {et~al.}(2017{\natexlab{b}})\citenamefont {Abbott} \emph
  {et~al.}}]{Abbott:2017vtc}%
  \BibitemOpen
  \bibfield  {author} {\bibinfo {author} {\bibfnamefont {B.~P.}\ \bibnamefont
  {Abbott}} \emph {et~al.} (\bibinfo {collaboration} {VIRGO, LIGO
  Scientific}),\ }\href {\doibase 10.1103/PhysRevLett.118.221101} {\bibfield
  {journal} {\bibinfo  {journal} {Phys. Rev. Lett.}\ }\textbf {\bibinfo
  {volume} {118}},\ \bibinfo {pages} {221101} (\bibinfo {year}
  {2017}{\natexlab{b}})},\ \Eprint {http://arxiv.org/abs/1706.01812}
  {arXiv:1706.01812 [gr-qc]} \BibitemShut {NoStop}%
\bibitem [{\citenamefont {Alexander}\ \emph {et~al.}(2008)\citenamefont
  {Alexander}, \citenamefont {Finn},\ and\ \citenamefont
  {Yunes}}]{Alexander:2007kv}%
  \BibitemOpen
  \bibfield  {author} {\bibinfo {author} {\bibfnamefont {S.}~\bibnamefont
  {Alexander}}, \bibinfo {author} {\bibfnamefont {L.~S.}\ \bibnamefont {Finn}},
  \ and\ \bibinfo {author} {\bibfnamefont {N.}~\bibnamefont {Yunes}},\ }\href
  {\doibase 10.1103/PhysRevD.78.066005} {\bibfield  {journal} {\bibinfo
  {journal} {Phys. Rev.}\ }\textbf {\bibinfo {volume} {D78}},\ \bibinfo {pages}
  {066005} (\bibinfo {year} {2008})},\ \Eprint {http://arxiv.org/abs/0712.2542}
  {arXiv:0712.2542 [gr-qc]} \BibitemShut {NoStop}%
\bibitem [{\citenamefont {Yunes}\ and\ \citenamefont
  {Finn}(2009)}]{Yunes:2008bu}%
  \BibitemOpen
  \bibfield  {author} {\bibinfo {author} {\bibfnamefont {N.}~\bibnamefont
  {Yunes}}\ and\ \bibinfo {author} {\bibfnamefont {L.~S.}\ \bibnamefont
  {Finn}},\ }\bibfield  {booktitle} {\emph {\bibinfo {booktitle} {{Laser
  Interferometer Space Antenna. Proceedings, 7th international LISA Symposium,
  Barcelona, Spain, June 16-20, 2008}}},\ }\href {\doibase
  10.1088/1742-6596/154/1/012041} {\bibfield  {journal} {\bibinfo  {journal}
  {J. Phys. Conf. Ser.}\ }\textbf {\bibinfo {volume} {154}},\ \bibinfo {pages}
  {012041} (\bibinfo {year} {2009})},\ \Eprint {http://arxiv.org/abs/0811.0181}
  {arXiv:0811.0181 [gr-qc]} \BibitemShut {NoStop}%
\bibitem [{\citenamefont {Yunes}\ \emph
  {et~al.}(2010{\natexlab{a}})\citenamefont {Yunes}, \citenamefont
  {O'Shaughnessy}, \citenamefont {Owen},\ and\ \citenamefont
  {Alexander}}]{Yunes:2010yf}%
  \BibitemOpen
  \bibfield  {author} {\bibinfo {author} {\bibfnamefont {N.}~\bibnamefont
  {Yunes}}, \bibinfo {author} {\bibfnamefont {R.}~\bibnamefont
  {O'Shaughnessy}}, \bibinfo {author} {\bibfnamefont {B.~J.}\ \bibnamefont
  {Owen}}, \ and\ \bibinfo {author} {\bibfnamefont {S.}~\bibnamefont
  {Alexander}},\ }\href {\doibase 10.1103/PhysRevD.82.064017} {\bibfield
  {journal} {\bibinfo  {journal} {Phys. Rev.}\ }\textbf {\bibinfo {volume}
  {D82}},\ \bibinfo {pages} {064017} (\bibinfo {year} {2010}{\natexlab{a}})},\
  \Eprint {http://arxiv.org/abs/1005.3310} {arXiv:1005.3310 [gr-qc]}
  \BibitemShut {NoStop}%
\bibitem [{\citenamefont {Yagi}\ and\ \citenamefont
  {Yang}(2018)}]{Yagi:2017zhb}%
  \BibitemOpen
  \bibfield  {author} {\bibinfo {author} {\bibfnamefont {K.}~\bibnamefont
  {Yagi}}\ and\ \bibinfo {author} {\bibfnamefont {H.}~\bibnamefont {Yang}},\
  }\href {\doibase 10.1103/PhysRevD.97.104018} {\bibfield  {journal} {\bibinfo
  {journal} {Phys. Rev.}\ }\textbf {\bibinfo {volume} {D97}},\ \bibinfo {pages}
  {104018} (\bibinfo {year} {2018})},\ \Eprint
  {http://arxiv.org/abs/1712.00682} {arXiv:1712.00682 [gr-qc]} \BibitemShut
  {NoStop}%
\bibitem [{\citenamefont {Maselli}\ \emph {et~al.}(2016)\citenamefont
  {Maselli}, \citenamefont {Marassi}, \citenamefont {Ferrari}, \citenamefont
  {Kokkotas},\ and\ \citenamefont {Schneider}}]{Maselli:2016ekw}%
  \BibitemOpen
  \bibfield  {author} {\bibinfo {author} {\bibfnamefont {A.}~\bibnamefont
  {Maselli}}, \bibinfo {author} {\bibfnamefont {S.}~\bibnamefont {Marassi}},
  \bibinfo {author} {\bibfnamefont {V.}~\bibnamefont {Ferrari}}, \bibinfo
  {author} {\bibfnamefont {K.}~\bibnamefont {Kokkotas}}, \ and\ \bibinfo
  {author} {\bibfnamefont {R.}~\bibnamefont {Schneider}},\ }\href {\doibase
  10.1103/PhysRevLett.117.091102} {\bibfield  {journal} {\bibinfo  {journal}
  {Phys. Rev. Lett.}\ }\textbf {\bibinfo {volume} {117}},\ \bibinfo {pages}
  {091102} (\bibinfo {year} {2016})},\ \Eprint
  {http://arxiv.org/abs/1606.04996} {arXiv:1606.04996 [gr-qc]} \BibitemShut
  {NoStop}%
\bibitem [{\citenamefont {Cardoso}\ \emph {et~al.}(2003)\citenamefont
  {Cardoso}, \citenamefont {Dias},\ and\ \citenamefont
  {Lemos}}]{Cardoso:2002pa}%
  \BibitemOpen
  \bibfield  {author} {\bibinfo {author} {\bibfnamefont {V.}~\bibnamefont
  {Cardoso}}, \bibinfo {author} {\bibfnamefont {O.~J.~C.}\ \bibnamefont
  {Dias}}, \ and\ \bibinfo {author} {\bibfnamefont {J.~P.~S.}\ \bibnamefont
  {Lemos}},\ }\href {\doibase 10.1103/PhysRevD.67.064026} {\bibfield  {journal}
  {\bibinfo  {journal} {Phys. Rev.}\ }\textbf {\bibinfo {volume} {D67}},\
  \bibinfo {pages} {064026} (\bibinfo {year} {2003})},\ \Eprint
  {http://arxiv.org/abs/hep-th/0212168} {arXiv:hep-th/0212168 [hep-th]}
  \BibitemShut {NoStop}%
\bibitem [{\citenamefont {Saltas}\ \emph {et~al.}(2014)\citenamefont {Saltas},
  \citenamefont {Sawicki}, \citenamefont {Amendola},\ and\ \citenamefont
  {Kunz}}]{Saltas:2014dha}%
  \BibitemOpen
  \bibfield  {author} {\bibinfo {author} {\bibfnamefont {I.~D.}\ \bibnamefont
  {Saltas}}, \bibinfo {author} {\bibfnamefont {I.}~\bibnamefont {Sawicki}},
  \bibinfo {author} {\bibfnamefont {L.}~\bibnamefont {Amendola}}, \ and\
  \bibinfo {author} {\bibfnamefont {M.}~\bibnamefont {Kunz}},\ }\href {\doibase
  10.1103/PhysRevLett.113.191101} {\bibfield  {journal} {\bibinfo  {journal}
  {Phys. Rev. Lett.}\ }\textbf {\bibinfo {volume} {113}},\ \bibinfo {pages}
  {191101} (\bibinfo {year} {2014})},\ \Eprint {http://arxiv.org/abs/1406.7139}
  {arXiv:1406.7139 [astro-ph.CO]} \BibitemShut {NoStop}%
\bibitem [{\citenamefont {Hwang}\ and\ \citenamefont
  {Noh}(1996)}]{Hwang:1996xh}%
  \BibitemOpen
  \bibfield  {author} {\bibinfo {author} {\bibfnamefont {J.-c.}\ \bibnamefont
  {Hwang}}\ and\ \bibinfo {author} {\bibfnamefont {H.}~\bibnamefont {Noh}},\
  }\href {\doibase 10.1103/PhysRevD.54.1460} {\bibfield  {journal} {\bibinfo
  {journal} {Phys. Rev.}\ }\textbf {\bibinfo {volume} {D54}},\ \bibinfo {pages}
  {1460} (\bibinfo {year} {1996})}\BibitemShut {NoStop}%
\bibitem [{\citenamefont {Nishizawa}(2018)}]{Nishizawa:2017nef}%
  \BibitemOpen
  \bibfield  {author} {\bibinfo {author} {\bibfnamefont {A.}~\bibnamefont
  {Nishizawa}},\ }\href {\doibase 10.1103/PhysRevD.97.104037} {\bibfield
  {journal} {\bibinfo  {journal} {Phys. Rev.}\ }\textbf {\bibinfo {volume}
  {D97}},\ \bibinfo {pages} {104037} (\bibinfo {year} {2018})},\ \Eprint
  {http://arxiv.org/abs/1710.04825} {arXiv:1710.04825 [gr-qc]} \BibitemShut
  {NoStop}%
\bibitem [{\citenamefont {Cornish}\ \emph {et~al.}(2011)\citenamefont
  {Cornish}, \citenamefont {Sampson}, \citenamefont {Yunes},\ and\
  \citenamefont {Pretorius}}]{Cornish:2011ys}%
  \BibitemOpen
  \bibfield  {author} {\bibinfo {author} {\bibfnamefont {N.}~\bibnamefont
  {Cornish}}, \bibinfo {author} {\bibfnamefont {L.}~\bibnamefont {Sampson}},
  \bibinfo {author} {\bibfnamefont {N.}~\bibnamefont {Yunes}}, \ and\ \bibinfo
  {author} {\bibfnamefont {F.}~\bibnamefont {Pretorius}},\ }\href {\doibase
  10.1103/PhysRevD.84.062003} {\bibfield  {journal} {\bibinfo  {journal} {Phys.
  Rev.}\ }\textbf {\bibinfo {volume} {D84}},\ \bibinfo {pages} {062003}
  (\bibinfo {year} {2011})},\ \Eprint {http://arxiv.org/abs/1105.2088}
  {arXiv:1105.2088 [gr-qc]} \BibitemShut {NoStop}%
\bibitem [{\citenamefont {Arun}(2012)}]{Arun:2012hf}%
  \BibitemOpen
  \bibfield  {author} {\bibinfo {author} {\bibfnamefont {K.~G.}\ \bibnamefont
  {Arun}},\ }\href {\doibase 10.1088/0264-9381/29/7/075011} {\bibfield
  {journal} {\bibinfo  {journal} {Class. Quant. Grav.}\ }\textbf {\bibinfo
  {volume} {29}},\ \bibinfo {pages} {075011} (\bibinfo {year} {2012})},\
  \Eprint {http://arxiv.org/abs/1202.5911} {arXiv:1202.5911 [gr-qc]}
  \BibitemShut {NoStop}%
\bibitem [{\citenamefont {Tahura}\ and\ \citenamefont
  {Yagi}(2018)}]{Tahura:2018zuq}%
  \BibitemOpen
  \bibfield  {author} {\bibinfo {author} {\bibfnamefont {S.}~\bibnamefont
  {Tahura}}\ and\ \bibinfo {author} {\bibfnamefont {K.}~\bibnamefont {Yagi}},\
  }\href {\doibase 10.1103/PhysRevD.98.084042} {\bibfield  {journal} {\bibinfo
  {journal} {Phys. Rev.}\ }\textbf {\bibinfo {volume} {D98}},\ \bibinfo {pages}
  {084042} (\bibinfo {year} {2018})},\ \Eprint
  {http://arxiv.org/abs/1809.00259} {arXiv:1809.00259 [gr-qc]} \BibitemShut
  {NoStop}%
\bibitem [{lig()}]{ligo:sample}%
  \BibitemOpen
  \href {https://dcc.ligo.org/LIGO-P1800370/public} {\enquote {\bibinfo {title}
  {Parameter estimation sample release for gwtc-1},}\ }\bibinfo {howpublished}
  {\url{https://dcc.ligo.org/LIGO-P1800370/public}},\ \bibinfo {note}
  {accessed: 2019-06-10}\BibitemShut {NoStop}%
\bibitem [{\citenamefont {Yagi}\ \emph {et~al.}(2012)\citenamefont {Yagi},
  \citenamefont {Stein}, \citenamefont {Yunes},\ and\ \citenamefont
  {Tanaka}}]{Yagi:2011xp}%
  \BibitemOpen
  \bibfield  {author} {\bibinfo {author} {\bibfnamefont {K.}~\bibnamefont
  {Yagi}}, \bibinfo {author} {\bibfnamefont {L.~C.}\ \bibnamefont {Stein}},
  \bibinfo {author} {\bibfnamefont {N.}~\bibnamefont {Yunes}}, \ and\ \bibinfo
  {author} {\bibfnamefont {T.}~\bibnamefont {Tanaka}},\ }\href {\doibase
  10.1103/PhysRevD.93.029902, 10.1103/PhysRevD.85.064022} {\bibfield  {journal}
  {\bibinfo  {journal} {Phys. Rev.}\ }\textbf {\bibinfo {volume} {D85}},\
  \bibinfo {pages} {064022} (\bibinfo {year} {2012})},\ \bibinfo {note}
  {[Erratum: Phys. Rev.D93,no.2,029902(2016)]},\ \Eprint
  {http://arxiv.org/abs/1110.5950} {arXiv:1110.5950 [gr-qc]} \BibitemShut
  {NoStop}%
\bibitem [{\citenamefont {Scharre}\ and\ \citenamefont
  {Will}(2002)}]{Scharre:2001hn}%
  \BibitemOpen
  \bibfield  {author} {\bibinfo {author} {\bibfnamefont {P.~D.}\ \bibnamefont
  {Scharre}}\ and\ \bibinfo {author} {\bibfnamefont {C.~M.}\ \bibnamefont
  {Will}},\ }\href {\doibase 10.1103/PhysRevD.65.042002} {\bibfield  {journal}
  {\bibinfo  {journal} {Phys. Rev.}\ }\textbf {\bibinfo {volume} {D65}},\
  \bibinfo {pages} {042002} (\bibinfo {year} {2002})},\ \Eprint
  {http://arxiv.org/abs/gr-qc/0109044} {arXiv:gr-qc/0109044 [gr-qc]}
  \BibitemShut {NoStop}%
\bibitem [{\citenamefont {Berti}\ \emph {et~al.}(2005)\citenamefont {Berti},
  \citenamefont {Buonanno},\ and\ \citenamefont {Will}}]{Berti:2004bd}%
  \BibitemOpen
  \bibfield  {author} {\bibinfo {author} {\bibfnamefont {E.}~\bibnamefont
  {Berti}}, \bibinfo {author} {\bibfnamefont {A.}~\bibnamefont {Buonanno}}, \
  and\ \bibinfo {author} {\bibfnamefont {C.~M.}\ \bibnamefont {Will}},\ }\href
  {\doibase 10.1103/PhysRevD.71.084025} {\bibfield  {journal} {\bibinfo
  {journal} {Phys. Rev.}\ }\textbf {\bibinfo {volume} {D71}},\ \bibinfo {pages}
  {084025} (\bibinfo {year} {2005})},\ \Eprint
  {http://arxiv.org/abs/gr-qc/0411129} {arXiv:gr-qc/0411129 [gr-qc]}
  \BibitemShut {NoStop}%
\bibitem [{\citenamefont {Yunes}\ \emph
  {et~al.}(2010{\natexlab{b}})\citenamefont {Yunes}, \citenamefont
  {Pretorius},\ and\ \citenamefont {Spergel}}]{Yunes:2009bv}%
  \BibitemOpen
  \bibfield  {author} {\bibinfo {author} {\bibfnamefont {N.}~\bibnamefont
  {Yunes}}, \bibinfo {author} {\bibfnamefont {F.}~\bibnamefont {Pretorius}}, \
  and\ \bibinfo {author} {\bibfnamefont {D.}~\bibnamefont {Spergel}},\ }\href
  {\doibase 10.1103/PhysRevD.81.064018} {\bibfield  {journal} {\bibinfo
  {journal} {Phys. Rev.}\ }\textbf {\bibinfo {volume} {D81}},\ \bibinfo {pages}
  {064018} (\bibinfo {year} {2010}{\natexlab{b}})},\ \Eprint
  {http://arxiv.org/abs/0912.2724} {arXiv:0912.2724 [gr-qc]} \BibitemShut
  {NoStop}%
\bibitem [{\citenamefont {{Cutler}}\ and\ \citenamefont
  {{Flanagan}}(1994)}]{cutlerflanagan}%
  \BibitemOpen
  \bibfield  {author} {\bibinfo {author} {\bibfnamefont {C.}~\bibnamefont
  {{Cutler}}}\ and\ \bibinfo {author} {\bibfnamefont {{\'E}.~E.}\ \bibnamefont
  {{Flanagan}}},\ }\href {\doibase 10.1103/PhysRevD.49.2658} {\bibfield
  {journal} {\bibinfo  {journal} {\prd}\ }\textbf {\bibinfo {volume} {49}},\
  \bibinfo {pages} {2658} (\bibinfo {year} {1994})},\ \Eprint
  {http://arxiv.org/abs/arXiv:gr-qc/9402014} {arXiv:gr-qc/9402014} \BibitemShut
  {NoStop}%
\bibitem [{\citenamefont {Blanchet}\ \emph {et~al.}(1995)\citenamefont
  {Blanchet}, \citenamefont {Damour}, \citenamefont {Iyer}, \citenamefont
  {Will},\ and\ \citenamefont {Wiseman}}]{Blanchet:1995ez}%
  \BibitemOpen
  \bibfield  {author} {\bibinfo {author} {\bibfnamefont {L.}~\bibnamefont
  {Blanchet}}, \bibinfo {author} {\bibfnamefont {T.}~\bibnamefont {Damour}},
  \bibinfo {author} {\bibfnamefont {B.~R.}\ \bibnamefont {Iyer}}, \bibinfo
  {author} {\bibfnamefont {C.~M.}\ \bibnamefont {Will}}, \ and\ \bibinfo
  {author} {\bibfnamefont {A.}~\bibnamefont {Wiseman}},\ }\href {\doibase
  10.1103/PhysRevLett.74.3515} {\bibfield  {journal} {\bibinfo  {journal}
  {Phys. Rev. Lett.}\ }\textbf {\bibinfo {volume} {74}},\ \bibinfo {pages}
  {3515} (\bibinfo {year} {1995})},\ \Eprint
  {http://arxiv.org/abs/gr-qc/9501027} {arXiv:gr-qc/9501027 [gr-qc]}
  \BibitemShut {NoStop}%
\bibitem [{\citenamefont {Damour}\ \emph {et~al.}(2000)\citenamefont {Damour},
  \citenamefont {Iyer},\ and\ \citenamefont
  {Sathyaprakash}}]{PhysRevD.62.084036}%
  \BibitemOpen
  \bibfield  {author} {\bibinfo {author} {\bibfnamefont {T.}~\bibnamefont
  {Damour}}, \bibinfo {author} {\bibfnamefont {B.~R.}\ \bibnamefont {Iyer}}, \
  and\ \bibinfo {author} {\bibfnamefont {B.~S.}\ \bibnamefont
  {Sathyaprakash}},\ }\href {\doibase 10.1103/PhysRevD.62.084036} {\bibfield
  {journal} {\bibinfo  {journal} {Phys. Rev. D}\ }\textbf {\bibinfo {volume}
  {62}},\ \bibinfo {pages} {084036} (\bibinfo {year} {2000})}\BibitemShut
  {NoStop}%
\bibitem [{\citenamefont {Yunes}\ \emph {et~al.}(2009)\citenamefont {Yunes},
  \citenamefont {Arun}, \citenamefont {Berti},\ and\ \citenamefont
  {Will}}]{Yunes:2009yz}%
  \BibitemOpen
  \bibfield  {author} {\bibinfo {author} {\bibfnamefont {N.}~\bibnamefont
  {Yunes}}, \bibinfo {author} {\bibfnamefont {K.~G.}\ \bibnamefont {Arun}},
  \bibinfo {author} {\bibfnamefont {E.}~\bibnamefont {Berti}}, \ and\ \bibinfo
  {author} {\bibfnamefont {C.~M.}\ \bibnamefont {Will}},\ }\href {\doibase
  10.1103/PhysRevD.89.109901, 10.1103/PhysRevD.80.084001} {\bibfield  {journal}
  {\bibinfo  {journal} {Phys. Rev.}\ }\textbf {\bibinfo {volume} {D80}},\
  \bibinfo {pages} {084001} (\bibinfo {year} {2009})},\ \bibinfo {note}
  {[Erratum: Phys. Rev.D89,no.10,109901(2014)]},\ \Eprint
  {http://arxiv.org/abs/0906.0313} {arXiv:0906.0313 [gr-qc]} \BibitemShut
  {NoStop}%
\bibitem [{\citenamefont {Blanchet}(2002)}]{Blanchet:2002av}%
  \BibitemOpen
  \bibfield  {author} {\bibinfo {author} {\bibfnamefont {L.}~\bibnamefont
  {Blanchet}},\ }\href {\doibase 10.12942/lrr-2002-3} {\bibfield  {journal}
  {\bibinfo  {journal} {Living Rev. Rel.}\ }\textbf {\bibinfo {volume} {5}},\
  \bibinfo {pages} {3} (\bibinfo {year} {2002})},\ \Eprint
  {http://arxiv.org/abs/gr-qc/0202016} {arXiv:gr-qc/0202016 [gr-qc]}
  \BibitemShut {NoStop}%
\bibitem [{\citenamefont {Nordtvedt}(1990)}]{PhysRevLett.65.953}%
  \BibitemOpen
  \bibfield  {author} {\bibinfo {author} {\bibfnamefont {K.}~\bibnamefont
  {Nordtvedt}},\ }\href {\doibase 10.1103/PhysRevLett.65.953} {\bibfield
  {journal} {\bibinfo  {journal} {Phys. Rev. Lett.}\ }\textbf {\bibinfo
  {volume} {65}},\ \bibinfo {pages} {953} (\bibinfo {year} {1990})}\BibitemShut
  {NoStop}%
\bibitem [{\citenamefont {Cutler}\ and\ \citenamefont
  {Flanagan}(1994)}]{Cutler:1994ys}%
  \BibitemOpen
  \bibfield  {author} {\bibinfo {author} {\bibfnamefont {C.}~\bibnamefont
  {Cutler}}\ and\ \bibinfo {author} {\bibfnamefont {E.~E.}\ \bibnamefont
  {Flanagan}},\ }\href {\doibase 10.1103/PhysRevD.49.2658} {\bibfield
  {journal} {\bibinfo  {journal} {Phys. Rev.}\ }\textbf {\bibinfo {volume}
  {D49}},\ \bibinfo {pages} {2658} (\bibinfo {year} {1994})},\ \Eprint
  {http://arxiv.org/abs/gr-qc/9402014} {arXiv:gr-qc/9402014 [gr-qc]}
  \BibitemShut {NoStop}%
\bibitem [{\citenamefont {Abbott}\ \emph
  {et~al.}(2018{\natexlab{b}})\citenamefont {Abbott} \emph
  {et~al.}}]{LIGOScientific:2018mvr}%
  \BibitemOpen
  \bibfield  {author} {\bibinfo {author} {\bibfnamefont {B.~P.}\ \bibnamefont
  {Abbott}} \emph {et~al.} (\bibinfo {collaboration} {LIGO Scientific,
  Virgo}),\ }\href@noop {} {\  (\bibinfo {year} {2018}{\natexlab{b}})},\
  \Eprint {http://arxiv.org/abs/1811.12907} {arXiv:1811.12907 [astro-ph.HE]}
  \BibitemShut {NoStop}%
\bibitem [{\citenamefont {Ajith}\ \emph {et~al.}(2011)\citenamefont {Ajith}
  \emph {et~al.}}]{Ajith:2009bn}%
  \BibitemOpen
  \bibfield  {author} {\bibinfo {author} {\bibfnamefont {P.}~\bibnamefont
  {Ajith}} \emph {et~al.},\ }\href {\doibase 10.1103/PhysRevLett.106.241101}
  {\bibfield  {journal} {\bibinfo  {journal} {Phys. Rev. Lett.}\ }\textbf
  {\bibinfo {volume} {106}},\ \bibinfo {pages} {241101} (\bibinfo {year}
  {2011})},\ \Eprint {http://arxiv.org/abs/0909.2867} {arXiv:0909.2867 [gr-qc]}
  \BibitemShut {NoStop}%
\bibitem [{\citenamefont {{R{\"o}ver}}\ and\ \citenamefont
  {{Friede}}(2016)}]{2016arXiv160204060R}%
  \BibitemOpen
  \bibfield  {author} {\bibinfo {author} {\bibfnamefont {C.}~\bibnamefont
  {{R{\"o}ver}}}\ and\ \bibinfo {author} {\bibfnamefont {T.}~\bibnamefont
  {{Friede}}},\ }\href@noop {} {\bibfield  {journal} {\bibinfo  {journal}
  {arXiv e-prints}\ ,\ \bibinfo {eid} {arXiv:1602.04060}} (\bibinfo {year}
  {2016})},\ \Eprint {http://arxiv.org/abs/1602.04060} {arXiv:1602.04060
  [stat.CO]} \BibitemShut {NoStop}%
\bibitem [{\citenamefont {Fierz}\ and\ \citenamefont
  {Pauli}(1939)}]{Fierz:1939ix}%
  \BibitemOpen
  \bibfield  {author} {\bibinfo {author} {\bibfnamefont {M.}~\bibnamefont
  {Fierz}}\ and\ \bibinfo {author} {\bibfnamefont {W.}~\bibnamefont {Pauli}},\
  }\href {\doibase 10.1098/rspa.1939.0140} {\bibfield  {journal} {\bibinfo
  {journal} {Proc. Roy. Soc. Lond.}\ }\textbf {\bibinfo {volume} {A173}},\
  \bibinfo {pages} {211} (\bibinfo {year} {1939})}\BibitemShut {NoStop}%
\bibitem [{\citenamefont {de~Rham}(2014)}]{deRham:2014zqa}%
  \BibitemOpen
  \bibfield  {author} {\bibinfo {author} {\bibfnamefont {C.}~\bibnamefont
  {de~Rham}},\ }\href {\doibase 10.12942/lrr-2014-7} {\bibfield  {journal}
  {\bibinfo  {journal} {Living Rev. Rel.}\ }\textbf {\bibinfo {volume} {17}},\
  \bibinfo {pages} {7} (\bibinfo {year} {2014})},\ \Eprint
  {http://arxiv.org/abs/1401.4173} {arXiv:1401.4173 [hep-th]} \BibitemShut
  {NoStop}%
\bibitem [{\citenamefont {Hinterbichler}(2012)}]{Hinterbichler:2011tt}%
  \BibitemOpen
  \bibfield  {author} {\bibinfo {author} {\bibfnamefont {K.}~\bibnamefont
  {Hinterbichler}},\ }\href {\doibase 10.1103/RevModPhys.84.671} {\bibfield
  {journal} {\bibinfo  {journal} {Rev. Mod. Phys.}\ }\textbf {\bibinfo {volume}
  {84}},\ \bibinfo {pages} {671} (\bibinfo {year} {2012})},\ \Eprint
  {http://arxiv.org/abs/1105.3735} {arXiv:1105.3735 [hep-th]} \BibitemShut
  {NoStop}%
\bibitem [{\citenamefont {de~Paula}\ \emph {et~al.}(2004)\citenamefont
  {de~Paula}, \citenamefont {Miranda},\ and\ \citenamefont
  {Marinho}}]{dePaula:2004bc}%
  \BibitemOpen
  \bibfield  {author} {\bibinfo {author} {\bibfnamefont {W.~L.~S.}\
  \bibnamefont {de~Paula}}, \bibinfo {author} {\bibfnamefont {O.~D.}\
  \bibnamefont {Miranda}}, \ and\ \bibinfo {author} {\bibfnamefont {R.~M.}\
  \bibnamefont {Marinho}},\ }\href {\doibase 10.1088/0264-9381/21/19/008}
  {\bibfield  {journal} {\bibinfo  {journal} {Class. Quant. Grav.}\ }\textbf
  {\bibinfo {volume} {21}},\ \bibinfo {pages} {4595} (\bibinfo {year}
  {2004})},\ \Eprint {http://arxiv.org/abs/gr-qc/0409041} {arXiv:gr-qc/0409041
  [gr-qc]} \BibitemShut {NoStop}%
\bibitem [{\citenamefont {Will}(1998)}]{Will:1997bb}%
  \BibitemOpen
  \bibfield  {author} {\bibinfo {author} {\bibfnamefont {C.~M.}\ \bibnamefont
  {Will}},\ }\href {\doibase 10.1103/PhysRevD.57.2061} {\bibfield  {journal}
  {\bibinfo  {journal} {Phys. Rev.}\ }\textbf {\bibinfo {volume} {D57}},\
  \bibinfo {pages} {2061} (\bibinfo {year} {1998})},\ \Eprint
  {http://arxiv.org/abs/gr-qc/9709011} {arXiv:gr-qc/9709011 [gr-qc]}
  \BibitemShut {NoStop}%
\bibitem [{\citenamefont {Finn}\ and\ \citenamefont
  {Sutton}(2002)}]{Finn:2001qi}%
  \BibitemOpen
  \bibfield  {author} {\bibinfo {author} {\bibfnamefont {L.~S.}\ \bibnamefont
  {Finn}}\ and\ \bibinfo {author} {\bibfnamefont {P.~J.}\ \bibnamefont
  {Sutton}},\ }\href {\doibase 10.1103/PhysRevD.65.044022} {\bibfield
  {journal} {\bibinfo  {journal} {Phys. Rev.}\ }\textbf {\bibinfo {volume}
  {D65}},\ \bibinfo {pages} {044022} (\bibinfo {year} {2002})},\ \Eprint
  {http://arxiv.org/abs/gr-qc/0109049} {arXiv:gr-qc/0109049 [gr-qc]}
  \BibitemShut {NoStop}%
\bibitem [{\citenamefont {Miao}\ \emph {et~al.}(2019)\citenamefont {Miao},
  \citenamefont {Shao},\ and\ \citenamefont {Ma}}]{Miao:2019nhf}%
  \BibitemOpen
  \bibfield  {author} {\bibinfo {author} {\bibfnamefont {X.}~\bibnamefont
  {Miao}}, \bibinfo {author} {\bibfnamefont {L.}~\bibnamefont {Shao}}, \ and\
  \bibinfo {author} {\bibfnamefont {B.-Q.}\ \bibnamefont {Ma}},\ }\href@noop {}
  {\  (\bibinfo {year} {2019})},\ \Eprint {http://arxiv.org/abs/1905.12836}
  {arXiv:1905.12836 [astro-ph.CO]} \BibitemShut {NoStop}%
\bibitem [{\citenamefont {Will}(2018)}]{Will:2018gku}%
  \BibitemOpen
  \bibfield  {author} {\bibinfo {author} {\bibfnamefont {C.~M.}\ \bibnamefont
  {Will}},\ }\href {\doibase 10.1088/1361-6382/aad13c} {\bibfield  {journal}
  {\bibinfo  {journal} {Class. Quant. Grav.}\ }\textbf {\bibinfo {volume}
  {35}},\ \bibinfo {pages} {17LT01} (\bibinfo {year} {2018})},\ \Eprint
  {http://arxiv.org/abs/1805.10523} {arXiv:1805.10523 [gr-qc]} \BibitemShut
  {NoStop}%
\bibitem [{\citenamefont {Goldhaber}\ and\ \citenamefont
  {Nieto}(1974)}]{Goldhaber:1974wg}%
  \BibitemOpen
  \bibfield  {author} {\bibinfo {author} {\bibfnamefont {A.~S.}\ \bibnamefont
  {Goldhaber}}\ and\ \bibinfo {author} {\bibfnamefont {M.~M.}\ \bibnamefont
  {Nieto}},\ }\href {\doibase 10.1103/PhysRevD.9.1119} {\bibfield  {journal}
  {\bibinfo  {journal} {Phys. Rev.}\ }\textbf {\bibinfo {volume} {D9}},\
  \bibinfo {pages} {1119} (\bibinfo {year} {1974})}\BibitemShut {NoStop}%
\bibitem [{\citenamefont {Gupta}\ and\ \citenamefont
  {Desai}(2018)}]{Gupta:2018hgm}%
  \BibitemOpen
  \bibfield  {author} {\bibinfo {author} {\bibfnamefont {S.}~\bibnamefont
  {Gupta}}\ and\ \bibinfo {author} {\bibfnamefont {S.}~\bibnamefont {Desai}},\
  }\href {\doibase 10.1016/j.aop.2018.09.017} {\bibfield  {journal} {\bibinfo
  {journal} {Annals Phys.}\ }\textbf {\bibinfo {volume} {399}},\ \bibinfo
  {pages} {85} (\bibinfo {year} {2018})},\ \Eprint
  {http://arxiv.org/abs/1810.00198} {arXiv:1810.00198 [astro-ph.CO]}
  \BibitemShut {NoStop}%
\bibitem [{\citenamefont {Desai}(2018)}]{Desai:2017dwg}%
  \BibitemOpen
  \bibfield  {author} {\bibinfo {author} {\bibfnamefont {S.}~\bibnamefont
  {Desai}},\ }\href {\doibase 10.1016/j.physletb.2018.01.052} {\bibfield
  {journal} {\bibinfo  {journal} {Phys. Lett.}\ }\textbf {\bibinfo {volume}
  {B778}},\ \bibinfo {pages} {325} (\bibinfo {year} {2018})},\ \Eprint
  {http://arxiv.org/abs/1708.06502} {arXiv:1708.06502 [astro-ph.CO]}
  \BibitemShut {NoStop}%
\bibitem [{\citenamefont {Choudhury}\ \emph {et~al.}(2004)\citenamefont
  {Choudhury}, \citenamefont {Joshi}, \citenamefont {Mahajan},\ and\
  \citenamefont {McKellar}}]{Choudhury:2002pu}%
  \BibitemOpen
  \bibfield  {author} {\bibinfo {author} {\bibfnamefont {S.~R.}\ \bibnamefont
  {Choudhury}}, \bibinfo {author} {\bibfnamefont {G.~C.}\ \bibnamefont
  {Joshi}}, \bibinfo {author} {\bibfnamefont {S.}~\bibnamefont {Mahajan}}, \
  and\ \bibinfo {author} {\bibfnamefont {B.~H.~J.}\ \bibnamefont {McKellar}},\
  }\href {\doibase 10.1016/j.astropartphys.2004.04.001} {\bibfield  {journal}
  {\bibinfo  {journal} {Astropart. Phys.}\ }\textbf {\bibinfo {volume} {21}},\
  \bibinfo {pages} {559} (\bibinfo {year} {2004})},\ \Eprint
  {http://arxiv.org/abs/hep-ph/0204161} {arXiv:hep-ph/0204161 [hep-ph]}
  \BibitemShut {NoStop}%
\bibitem [{\citenamefont {Brito}\ \emph {et~al.}(2013)\citenamefont {Brito},
  \citenamefont {Cardoso},\ and\ \citenamefont {Pani}}]{Brito:2013wya}%
  \BibitemOpen
  \bibfield  {author} {\bibinfo {author} {\bibfnamefont {R.}~\bibnamefont
  {Brito}}, \bibinfo {author} {\bibfnamefont {V.}~\bibnamefont {Cardoso}}, \
  and\ \bibinfo {author} {\bibfnamefont {P.}~\bibnamefont {Pani}},\ }\href
  {\doibase 10.1103/PhysRevD.88.023514} {\bibfield  {journal} {\bibinfo
  {journal} {Phys. Rev.}\ }\textbf {\bibinfo {volume} {D88}},\ \bibinfo {pages}
  {023514} (\bibinfo {year} {2013})},\ \Eprint {http://arxiv.org/abs/1304.6725}
  {arXiv:1304.6725 [gr-qc]} \BibitemShut {NoStop}%
\bibitem [{\citenamefont {Moura}\ and\ \citenamefont
  {Schiappa}(2007)}]{Moura:2006pz}%
  \BibitemOpen
  \bibfield  {author} {\bibinfo {author} {\bibfnamefont {F.}~\bibnamefont
  {Moura}}\ and\ \bibinfo {author} {\bibfnamefont {R.}~\bibnamefont
  {Schiappa}},\ }\href {\doibase 10.1088/0264-9381/24/2/006} {\bibfield
  {journal} {\bibinfo  {journal} {Class. Quant. Grav.}\ }\textbf {\bibinfo
  {volume} {24}},\ \bibinfo {pages} {361} (\bibinfo {year} {2007})},\ \Eprint
  {http://arxiv.org/abs/hep-th/0605001} {arXiv:hep-th/0605001 [hep-th]}
  \BibitemShut {NoStop}%
\bibitem [{\citenamefont {Pani}\ and\ \citenamefont
  {Cardoso}(2009)}]{Pani:2009wy}%
  \BibitemOpen
  \bibfield  {author} {\bibinfo {author} {\bibfnamefont {P.}~\bibnamefont
  {Pani}}\ and\ \bibinfo {author} {\bibfnamefont {V.}~\bibnamefont {Cardoso}},\
  }\href {\doibase 10.1103/PhysRevD.79.084031} {\bibfield  {journal} {\bibinfo
  {journal} {Phys. Rev.}\ }\textbf {\bibinfo {volume} {D79}},\ \bibinfo {pages}
  {084031} (\bibinfo {year} {2009})},\ \Eprint {http://arxiv.org/abs/0902.1569}
  {arXiv:0902.1569 [gr-qc]} \BibitemShut {NoStop}%
\bibitem [{\citenamefont {Zhang}\ \emph {et~al.}(2017)\citenamefont {Zhang},
  \citenamefont {Zhou}, \citenamefont {Bambi}, \citenamefont {Kleihaus},
  \citenamefont {Kunz},\ and\ \citenamefont {Radu}}]{Zhang:2017unx}%
  \BibitemOpen
  \bibfield  {author} {\bibinfo {author} {\bibfnamefont {H.}~\bibnamefont
  {Zhang}}, \bibinfo {author} {\bibfnamefont {M.}~\bibnamefont {Zhou}},
  \bibinfo {author} {\bibfnamefont {C.}~\bibnamefont {Bambi}}, \bibinfo
  {author} {\bibfnamefont {B.}~\bibnamefont {Kleihaus}}, \bibinfo {author}
  {\bibfnamefont {J.}~\bibnamefont {Kunz}}, \ and\ \bibinfo {author}
  {\bibfnamefont {E.}~\bibnamefont {Radu}},\ }\href {\doibase
  10.1103/PhysRevD.95.104043} {\bibfield  {journal} {\bibinfo  {journal} {Phys.
  Rev.}\ }\textbf {\bibinfo {volume} {D95}},\ \bibinfo {pages} {104043}
  (\bibinfo {year} {2017})},\ \Eprint {http://arxiv.org/abs/1704.04426}
  {arXiv:1704.04426 [gr-qc]} \BibitemShut {NoStop}%
\bibitem [{\citenamefont {Berti}\ \emph {et~al.}(2015)\citenamefont {Berti}
  \emph {et~al.}}]{Berti:2015itd}%
  \BibitemOpen
  \bibfield  {author} {\bibinfo {author} {\bibfnamefont {E.}~\bibnamefont
  {Berti}} \emph {et~al.},\ }\href {\doibase 10.1088/0264-9381/32/24/243001}
  {\bibfield  {journal} {\bibinfo  {journal} {Class. Quant. Grav.}\ }\textbf
  {\bibinfo {volume} {32}},\ \bibinfo {pages} {243001} (\bibinfo {year}
  {2015})},\ \Eprint {http://arxiv.org/abs/1501.07274} {arXiv:1501.07274
  [gr-qc]} \BibitemShut {NoStop}%
\bibitem [{\citenamefont {Kanti}\ \emph {et~al.}(1996)\citenamefont {Kanti},
  \citenamefont {Mavromatos}, \citenamefont {Rizos}, \citenamefont {Tamvakis},\
  and\ \citenamefont {Winstanley}}]{Kanti:1995vq}%
  \BibitemOpen
  \bibfield  {author} {\bibinfo {author} {\bibfnamefont {P.}~\bibnamefont
  {Kanti}}, \bibinfo {author} {\bibfnamefont {N.~E.}\ \bibnamefont
  {Mavromatos}}, \bibinfo {author} {\bibfnamefont {J.}~\bibnamefont {Rizos}},
  \bibinfo {author} {\bibfnamefont {K.}~\bibnamefont {Tamvakis}}, \ and\
  \bibinfo {author} {\bibfnamefont {E.}~\bibnamefont {Winstanley}},\ }\href
  {\doibase 10.1103/PhysRevD.54.5049} {\bibfield  {journal} {\bibinfo
  {journal} {Phys. Rev.}\ }\textbf {\bibinfo {volume} {D54}},\ \bibinfo {pages}
  {5049} (\bibinfo {year} {1996})},\ \Eprint
  {http://arxiv.org/abs/hep-th/9511071} {arXiv:hep-th/9511071 [hep-th]}
  \BibitemShut {NoStop}%
\bibitem [{\citenamefont {Sotiriou}\ and\ \citenamefont
  {Zhou}(2014)}]{Sotiriou:2014pfa}%
  \BibitemOpen
  \bibfield  {author} {\bibinfo {author} {\bibfnamefont {T.~P.}\ \bibnamefont
  {Sotiriou}}\ and\ \bibinfo {author} {\bibfnamefont {S.-Y.}\ \bibnamefont
  {Zhou}},\ }\href {\doibase 10.1103/PhysRevD.90.124063} {\bibfield  {journal}
  {\bibinfo  {journal} {Phys. Rev.}\ }\textbf {\bibinfo {volume} {D90}},\
  \bibinfo {pages} {124063} (\bibinfo {year} {2014})},\ \Eprint
  {http://arxiv.org/abs/1408.1698} {arXiv:1408.1698 [gr-qc]} \BibitemShut
  {NoStop}%
\bibitem [{\citenamefont {Berti}\ \emph {et~al.}(2018)\citenamefont {Berti},
  \citenamefont {Yagi},\ and\ \citenamefont {Yunes}}]{Berti:2018cxi}%
  \BibitemOpen
  \bibfield  {author} {\bibinfo {author} {\bibfnamefont {E.}~\bibnamefont
  {Berti}}, \bibinfo {author} {\bibfnamefont {K.}~\bibnamefont {Yagi}}, \ and\
  \bibinfo {author} {\bibfnamefont {N.}~\bibnamefont {Yunes}},\ }\href
  {\doibase 10.1007/s10714-018-2362-8} {\bibfield  {journal} {\bibinfo
  {journal} {Gen. Rel. Grav.}\ }\textbf {\bibinfo {volume} {50}},\ \bibinfo
  {pages} {46} (\bibinfo {year} {2018})},\ \Eprint
  {http://arxiv.org/abs/1801.03208} {arXiv:1801.03208 [gr-qc]} \BibitemShut
  {NoStop}%
\bibitem [{\citenamefont {Prabhu}\ and\ \citenamefont
  {Stein}(2018)}]{Prabhu:2018aun}%
  \BibitemOpen
  \bibfield  {author} {\bibinfo {author} {\bibfnamefont {K.}~\bibnamefont
  {Prabhu}}\ and\ \bibinfo {author} {\bibfnamefont {L.~C.}\ \bibnamefont
  {Stein}},\ }\href {\doibase 10.1103/PhysRevD.98.021503} {\bibfield  {journal}
  {\bibinfo  {journal} {Phys. Rev.}\ }\textbf {\bibinfo {volume} {D98}},\
  \bibinfo {pages} {021503} (\bibinfo {year} {2018})},\ \Eprint
  {http://arxiv.org/abs/1805.02668} {arXiv:1805.02668 [gr-qc]} \BibitemShut
  {NoStop}%
\bibitem [{\citenamefont {Yagi}\ \emph {et~al.}(2016)\citenamefont {Yagi},
  \citenamefont {Stein},\ and\ \citenamefont {Yunes}}]{Yagi:2015oca}%
  \BibitemOpen
  \bibfield  {author} {\bibinfo {author} {\bibfnamefont {K.}~\bibnamefont
  {Yagi}}, \bibinfo {author} {\bibfnamefont {L.~C.}\ \bibnamefont {Stein}}, \
  and\ \bibinfo {author} {\bibfnamefont {N.}~\bibnamefont {Yunes}},\ }\href
  {\doibase 10.1103/PhysRevD.93.024010} {\bibfield  {journal} {\bibinfo
  {journal} {Phys. Rev.}\ }\textbf {\bibinfo {volume} {D93}},\ \bibinfo {pages}
  {024010} (\bibinfo {year} {2016})},\ \Eprint
  {http://arxiv.org/abs/1510.02152} {arXiv:1510.02152 [gr-qc]} \BibitemShut
  {NoStop}%
\bibitem [{\citenamefont {Nair}\ \emph {et~al.}(2019)\citenamefont {Nair},
  \citenamefont {Perkins}, \citenamefont {Silva},\ and\ \citenamefont
  {Yunes}}]{Nair:2019iur}%
  \BibitemOpen
  \bibfield  {author} {\bibinfo {author} {\bibfnamefont {R.}~\bibnamefont
  {Nair}}, \bibinfo {author} {\bibfnamefont {S.}~\bibnamefont {Perkins}},
  \bibinfo {author} {\bibfnamefont {H.~O.}\ \bibnamefont {Silva}}, \ and\
  \bibinfo {author} {\bibfnamefont {N.}~\bibnamefont {Yunes}},\ }\href@noop {}
  {\  (\bibinfo {year} {2019})},\ \Eprint {http://arxiv.org/abs/1905.00870}
  {arXiv:1905.00870 [gr-qc]} \BibitemShut {NoStop}%
\bibitem [{\citenamefont {Yamada}\ \emph {et~al.}(2019)\citenamefont {Yamada},
  \citenamefont {Narikawa},\ and\ \citenamefont {Tanaka}}]{Yamada:2019zrb}%
  \BibitemOpen
  \bibfield  {author} {\bibinfo {author} {\bibfnamefont {K.}~\bibnamefont
  {Yamada}}, \bibinfo {author} {\bibfnamefont {T.}~\bibnamefont {Narikawa}}, \
  and\ \bibinfo {author} {\bibfnamefont {T.}~\bibnamefont {Tanaka}},\
  }\href@noop {} {\  (\bibinfo {year} {2019})},\ \Eprint
  {http://arxiv.org/abs/1905.11859} {arXiv:1905.11859 [gr-qc]} \BibitemShut
  {NoStop}%
\bibitem [{\citenamefont {Yagi}(2012)}]{Yagi:2012gp}%
  \BibitemOpen
  \bibfield  {author} {\bibinfo {author} {\bibfnamefont {K.}~\bibnamefont
  {Yagi}},\ }\href {\doibase 10.1103/PhysRevD.86.081504} {\bibfield  {journal}
  {\bibinfo  {journal} {Phys. Rev.}\ }\textbf {\bibinfo {volume} {D86}},\
  \bibinfo {pages} {081504} (\bibinfo {year} {2012})},\ \Eprint
  {http://arxiv.org/abs/1204.4524} {arXiv:1204.4524 [gr-qc]} \BibitemShut
  {NoStop}%
\bibitem [{\citenamefont {Fujii}\ and\ \citenamefont
  {Maeda}(2007)}]{Fujii:2003pa}%
  \BibitemOpen
  \bibfield  {author} {\bibinfo {author} {\bibfnamefont {Y.}~\bibnamefont
  {Fujii}}\ and\ \bibinfo {author} {\bibfnamefont {K.}~\bibnamefont {Maeda}},\
  }\href {\doibase 10.1017/CBO9780511535093} {\emph {\bibinfo {title} {{The
  scalar-tensor theory of gravitation}}}},\ Cambridge Monographs on
  Mathematical Physics\ (\bibinfo  {publisher} {Cambridge University Press},\
  \bibinfo {year} {2007})\BibitemShut {NoStop}%
\bibitem [{\citenamefont {Overduin}\ and\ \citenamefont
  {Wesson}(1997)}]{Overduin:1998pn}%
  \BibitemOpen
  \bibfield  {author} {\bibinfo {author} {\bibfnamefont {J.~M.}\ \bibnamefont
  {Overduin}}\ and\ \bibinfo {author} {\bibfnamefont {P.~S.}\ \bibnamefont
  {Wesson}},\ }\href {\doibase 10.1016/S0370-1573(96)00046-4} {\bibfield
  {journal} {\bibinfo  {journal} {Phys. Rept.}\ }\textbf {\bibinfo {volume}
  {283}},\ \bibinfo {pages} {303} (\bibinfo {year} {1997})},\ \Eprint
  {http://arxiv.org/abs/gr-qc/9805018} {arXiv:gr-qc/9805018 [gr-qc]}
  \BibitemShut {NoStop}%
\bibitem [{\citenamefont {Polchinski}(1998{\natexlab{a}})}]{polchinski1}%
  \BibitemOpen
  \bibfield  {author} {\bibinfo {author} {\bibfnamefont {J.}~\bibnamefont
  {Polchinski}},\ }\href@noop {} {\emph {\bibinfo {title} {String theory. Vol.
  1: An introduction to the bosonic string}}}\ (\bibinfo  {publisher}
  {Cambridge University Press},\ \bibinfo {address} {Cambridge, UK},\ \bibinfo
  {year} {1998})\BibitemShut {NoStop}%
\bibitem [{\citenamefont {Polchinski}(1998{\natexlab{b}})}]{polchinski2}%
  \BibitemOpen
  \bibfield  {author} {\bibinfo {author} {\bibfnamefont {J.}~\bibnamefont
  {Polchinski}},\ }\href@noop {} {\emph {\bibinfo {title} {String theory. Vol.
  2: Superstring theory and beyond}}}\ (\bibinfo  {publisher} {Cambridge
  University Press},\ \bibinfo {address} {Cambridge, UK},\ \bibinfo {year}
  {1998})\BibitemShut {NoStop}%
\bibitem [{\citenamefont {Chiba}\ \emph {et~al.}(1997)\citenamefont {Chiba},
  \citenamefont {Harada},\ and\ \citenamefont {Nakao}}]{Chiba:1997ms}%
  \BibitemOpen
  \bibfield  {author} {\bibinfo {author} {\bibfnamefont {T.}~\bibnamefont
  {Chiba}}, \bibinfo {author} {\bibfnamefont {T.}~\bibnamefont {Harada}}, \
  and\ \bibinfo {author} {\bibfnamefont {K.-i.}\ \bibnamefont {Nakao}},\ }\href
  {\doibase 10.1143/PTPS.128.335} {\bibfield  {journal} {\bibinfo  {journal}
  {Prog. Theor. Phys. Suppl.}\ }\textbf {\bibinfo {volume} {128}},\ \bibinfo
  {pages} {335} (\bibinfo {year} {1997})}\BibitemShut {NoStop}%
\bibitem [{\citenamefont {Harrison}(1972)}]{PhysRevD.6.2077}%
  \BibitemOpen
  \bibfield  {author} {\bibinfo {author} {\bibfnamefont {E.~R.}\ \bibnamefont
  {Harrison}},\ }\href {\doibase 10.1103/PhysRevD.6.2077} {\bibfield  {journal}
  {\bibinfo  {journal} {Phys. Rev. D}\ }\textbf {\bibinfo {volume} {6}},\
  \bibinfo {pages} {2077} (\bibinfo {year} {1972})}\BibitemShut {NoStop}%
\bibitem [{\citenamefont {Brax}\ \emph {et~al.}(2004)\citenamefont {Brax},
  \citenamefont {van~de Bruck}, \citenamefont {Davis}, \citenamefont {Khoury},\
  and\ \citenamefont {Weltman}}]{Brax:2004qh}%
  \BibitemOpen
  \bibfield  {author} {\bibinfo {author} {\bibfnamefont {P.}~\bibnamefont
  {Brax}}, \bibinfo {author} {\bibfnamefont {C.}~\bibnamefont {van~de Bruck}},
  \bibinfo {author} {\bibfnamefont {A.-C.}\ \bibnamefont {Davis}}, \bibinfo
  {author} {\bibfnamefont {J.}~\bibnamefont {Khoury}}, \ and\ \bibinfo {author}
  {\bibfnamefont {A.}~\bibnamefont {Weltman}},\ }\href {\doibase
  10.1103/PhysRevD.70.123518} {\bibfield  {journal} {\bibinfo  {journal} {Phys.
  Rev.}\ }\textbf {\bibinfo {volume} {D70}},\ \bibinfo {pages} {123518}
  (\bibinfo {year} {2004})},\ \Eprint {http://arxiv.org/abs/astro-ph/0408415}
  {arXiv:astro-ph/0408415 [astro-ph]} \BibitemShut {NoStop}%
\bibitem [{\citenamefont {Kainulainen}\ and\ \citenamefont
  {Sunhede}(2006)}]{PhysRevD.73.083510}%
  \BibitemOpen
  \bibfield  {author} {\bibinfo {author} {\bibfnamefont {K.}~\bibnamefont
  {Kainulainen}}\ and\ \bibinfo {author} {\bibfnamefont {D.}~\bibnamefont
  {Sunhede}},\ }\href {\doibase 10.1103/PhysRevD.73.083510} {\bibfield
  {journal} {\bibinfo  {journal} {Phys. Rev. D}\ }\textbf {\bibinfo {volume}
  {73}},\ \bibinfo {pages} {083510} (\bibinfo {year} {2006})}\BibitemShut
  {NoStop}%
\bibitem [{\citenamefont {Baccigalupi}\ \emph {et~al.}(2000)\citenamefont
  {Baccigalupi}, \citenamefont {Matarrese},\ and\ \citenamefont
  {Perrotta}}]{PhysRevD.62.123510}%
  \BibitemOpen
  \bibfield  {author} {\bibinfo {author} {\bibfnamefont {C.}~\bibnamefont
  {Baccigalupi}}, \bibinfo {author} {\bibfnamefont {S.}~\bibnamefont
  {Matarrese}}, \ and\ \bibinfo {author} {\bibfnamefont {F.}~\bibnamefont
  {Perrotta}},\ }\href {\doibase 10.1103/PhysRevD.62.123510} {\bibfield
  {journal} {\bibinfo  {journal} {Phys. Rev. D}\ }\textbf {\bibinfo {volume}
  {62}},\ \bibinfo {pages} {123510} (\bibinfo {year} {2000})}\BibitemShut
  {NoStop}%
\bibitem [{\citenamefont {Riazuelo}\ and\ \citenamefont
  {Uzan}(2002)}]{PhysRevD.66.023525}%
  \BibitemOpen
  \bibfield  {author} {\bibinfo {author} {\bibfnamefont {A.}~\bibnamefont
  {Riazuelo}}\ and\ \bibinfo {author} {\bibfnamefont {J.-P.}\ \bibnamefont
  {Uzan}},\ }\href {\doibase 10.1103/PhysRevD.66.023525} {\bibfield  {journal}
  {\bibinfo  {journal} {Phys. Rev. D}\ }\textbf {\bibinfo {volume} {66}},\
  \bibinfo {pages} {023525} (\bibinfo {year} {2002})}\BibitemShut {NoStop}%
\bibitem [{\citenamefont {Schimd}\ \emph {et~al.}(2005)\citenamefont {Schimd},
  \citenamefont {Uzan},\ and\ \citenamefont {Riazuelo}}]{Schimd:2004nq}%
  \BibitemOpen
  \bibfield  {author} {\bibinfo {author} {\bibfnamefont {C.}~\bibnamefont
  {Schimd}}, \bibinfo {author} {\bibfnamefont {J.-P.}\ \bibnamefont {Uzan}}, \
  and\ \bibinfo {author} {\bibfnamefont {A.}~\bibnamefont {Riazuelo}},\ }\href
  {\doibase 10.1103/PhysRevD.71.083512} {\bibfield  {journal} {\bibinfo
  {journal} {Phys. Rev.}\ }\textbf {\bibinfo {volume} {D71}},\ \bibinfo {pages}
  {083512} (\bibinfo {year} {2005})},\ \Eprint
  {http://arxiv.org/abs/astro-ph/0412120} {arXiv:astro-ph/0412120 [astro-ph]}
  \BibitemShut {NoStop}%
\bibitem [{\citenamefont {Burd}\ and\ \citenamefont
  {Coley}(1991)}]{Burd:1991ns}%
  \BibitemOpen
  \bibfield  {author} {\bibinfo {author} {\bibfnamefont {A.}~\bibnamefont
  {Burd}}\ and\ \bibinfo {author} {\bibfnamefont {A.}~\bibnamefont {Coley}},\
  }\href {\doibase 10.1016/0370-2693(91)90941-I} {\bibfield  {journal}
  {\bibinfo  {journal} {Phys. Lett.}\ }\textbf {\bibinfo {volume} {B267}},\
  \bibinfo {pages} {330} (\bibinfo {year} {1991})}\BibitemShut {NoStop}%
\bibitem [{\citenamefont {Barrow}\ and\ \citenamefont
  {Maeda}(1990)}]{Barrow:1990nv}%
  \BibitemOpen
  \bibfield  {author} {\bibinfo {author} {\bibfnamefont {J.~D.}\ \bibnamefont
  {Barrow}}\ and\ \bibinfo {author} {\bibfnamefont {K.-i.}\ \bibnamefont
  {Maeda}},\ }\href {\doibase 10.1016/0550-3213(90)90272-F} {\bibfield
  {journal} {\bibinfo  {journal} {Nucl. Phys.}\ }\textbf {\bibinfo {volume}
  {B341}},\ \bibinfo {pages} {294} (\bibinfo {year} {1990})}\BibitemShut
  {NoStop}%
\bibitem [{\citenamefont {Clifton}\ \emph {et~al.}(2012)\citenamefont
  {Clifton}, \citenamefont {Ferreira}, \citenamefont {Padilla},\ and\
  \citenamefont {Skordis}}]{Clifton:2011jh}%
  \BibitemOpen
  \bibfield  {author} {\bibinfo {author} {\bibfnamefont {T.}~\bibnamefont
  {Clifton}}, \bibinfo {author} {\bibfnamefont {P.~G.}\ \bibnamefont
  {Ferreira}}, \bibinfo {author} {\bibfnamefont {A.}~\bibnamefont {Padilla}}, \
  and\ \bibinfo {author} {\bibfnamefont {C.}~\bibnamefont {Skordis}},\ }\href
  {\doibase 10.1016/j.physrep.2012.01.001} {\bibfield  {journal} {\bibinfo
  {journal} {Phys. Rept.}\ }\textbf {\bibinfo {volume} {513}},\ \bibinfo
  {pages} {1} (\bibinfo {year} {2012})},\ \Eprint
  {http://arxiv.org/abs/1106.2476} {arXiv:1106.2476 [astro-ph.CO]} \BibitemShut
  {NoStop}%
\bibitem [{\citenamefont {Coc}\ \emph {et~al.}(2006)\citenamefont {Coc},
  \citenamefont {Olive}, \citenamefont {Uzan},\ and\ \citenamefont
  {Vangioni}}]{Coc:2006rt}%
  \BibitemOpen
  \bibfield  {author} {\bibinfo {author} {\bibfnamefont {A.}~\bibnamefont
  {Coc}}, \bibinfo {author} {\bibfnamefont {K.~A.}\ \bibnamefont {Olive}},
  \bibinfo {author} {\bibfnamefont {J.-P.}\ \bibnamefont {Uzan}}, \ and\
  \bibinfo {author} {\bibfnamefont {E.}~\bibnamefont {Vangioni}},\ }\href
  {\doibase 10.1103/PhysRevD.73.083525} {\bibfield  {journal} {\bibinfo
  {journal} {Phys. Rev.}\ }\textbf {\bibinfo {volume} {D73}},\ \bibinfo {pages}
  {083525} (\bibinfo {year} {2006})},\ \Eprint
  {http://arxiv.org/abs/astro-ph/0601299} {arXiv:astro-ph/0601299 [astro-ph]}
  \BibitemShut {NoStop}%
\bibitem [{\citenamefont {Damour}\ and\ \citenamefont
  {Pichon}(1999)}]{Damour:1998ae}%
  \BibitemOpen
  \bibfield  {author} {\bibinfo {author} {\bibfnamefont {T.}~\bibnamefont
  {Damour}}\ and\ \bibinfo {author} {\bibfnamefont {B.}~\bibnamefont
  {Pichon}},\ }\href {\doibase 10.1103/PhysRevD.59.123502} {\bibfield
  {journal} {\bibinfo  {journal} {Phys. Rev.}\ }\textbf {\bibinfo {volume}
  {D59}},\ \bibinfo {pages} {123502} (\bibinfo {year} {1999})},\ \Eprint
  {http://arxiv.org/abs/astro-ph/9807176} {arXiv:astro-ph/9807176 [astro-ph]}
  \BibitemShut {NoStop}%
\bibitem [{\citenamefont {Larena}\ \emph {et~al.}(2007)\citenamefont {Larena},
  \citenamefont {Alimi},\ and\ \citenamefont {Serna}}]{Larena:2005tu}%
  \BibitemOpen
  \bibfield  {author} {\bibinfo {author} {\bibfnamefont {J.}~\bibnamefont
  {Larena}}, \bibinfo {author} {\bibfnamefont {J.-M.}\ \bibnamefont {Alimi}}, \
  and\ \bibinfo {author} {\bibfnamefont {A.}~\bibnamefont {Serna}},\ }\href
  {\doibase 10.1086/511028} {\bibfield  {journal} {\bibinfo  {journal}
  {Astrophys. J.}\ }\textbf {\bibinfo {volume} {658}},\ \bibinfo {pages} {1}
  (\bibinfo {year} {2007})},\ \Eprint {http://arxiv.org/abs/astro-ph/0511693}
  {arXiv:astro-ph/0511693 [astro-ph]} \BibitemShut {NoStop}%
\bibitem [{\citenamefont {Torres}(1995)}]{Torres:1995je}%
  \BibitemOpen
  \bibfield  {author} {\bibinfo {author} {\bibfnamefont {D.~F.}\ \bibnamefont
  {Torres}},\ }\href {\doibase 10.1016/0370-2693(95)01098-B} {\bibfield
  {journal} {\bibinfo  {journal} {Phys. Lett.}\ }\textbf {\bibinfo {volume}
  {B359}},\ \bibinfo {pages} {249} (\bibinfo {year} {1995})}\BibitemShut
  {NoStop}%
\bibitem [{\citenamefont {Brax}\ \emph {et~al.}(2006)\citenamefont {Brax},
  \citenamefont {van~de Bruck}, \citenamefont {Davis},\ and\ \citenamefont
  {Green}}]{Brax:2005ew}%
  \BibitemOpen
  \bibfield  {author} {\bibinfo {author} {\bibfnamefont {P.}~\bibnamefont
  {Brax}}, \bibinfo {author} {\bibfnamefont {C.}~\bibnamefont {van~de Bruck}},
  \bibinfo {author} {\bibfnamefont {A.-C.}\ \bibnamefont {Davis}}, \ and\
  \bibinfo {author} {\bibfnamefont {A.~M.}\ \bibnamefont {Green}},\ }\href
  {\doibase 10.1016/j.physletb.2005.12.055} {\bibfield  {journal} {\bibinfo
  {journal} {Phys. Lett.}\ }\textbf {\bibinfo {volume} {B633}},\ \bibinfo
  {pages} {441} (\bibinfo {year} {2006})},\ \Eprint
  {http://arxiv.org/abs/astro-ph/0509878} {arXiv:astro-ph/0509878 [astro-ph]}
  \BibitemShut {NoStop}%
\bibitem [{\citenamefont {Damour}\ and\ \citenamefont
  {Esposito-Far\`ese}(1993)}]{PhysRevLett.70.2220}%
  \BibitemOpen
  \bibfield  {author} {\bibinfo {author} {\bibfnamefont {T.}~\bibnamefont
  {Damour}}\ and\ \bibinfo {author} {\bibfnamefont {G.}~\bibnamefont
  {Esposito-Far\`ese}},\ }\href {\doibase 10.1103/PhysRevLett.70.2220}
  {\bibfield  {journal} {\bibinfo  {journal} {Phys. Rev. Lett.}\ }\textbf
  {\bibinfo {volume} {70}},\ \bibinfo {pages} {2220} (\bibinfo {year}
  {1993})}\BibitemShut {NoStop}%
\bibitem [{\citenamefont {Barausse}\ \emph {et~al.}(2013)\citenamefont
  {Barausse}, \citenamefont {Palenzuela}, \citenamefont {Ponce},\ and\
  \citenamefont {Lehner}}]{Barausse:2012da}%
  \BibitemOpen
  \bibfield  {author} {\bibinfo {author} {\bibfnamefont {E.}~\bibnamefont
  {Barausse}}, \bibinfo {author} {\bibfnamefont {C.}~\bibnamefont
  {Palenzuela}}, \bibinfo {author} {\bibfnamefont {M.}~\bibnamefont {Ponce}}, \
  and\ \bibinfo {author} {\bibfnamefont {L.}~\bibnamefont {Lehner}},\ }\href
  {\doibase 10.1103/PhysRevD.87.081506} {\bibfield  {journal} {\bibinfo
  {journal} {Phys. Rev.}\ }\textbf {\bibinfo {volume} {D87}},\ \bibinfo {pages}
  {081506} (\bibinfo {year} {2013})},\ \Eprint {http://arxiv.org/abs/1212.5053}
  {arXiv:1212.5053 [gr-qc]} \BibitemShut {NoStop}%
\bibitem [{\citenamefont {Jacobson}(1999)}]{Jacobson:1999vr}%
  \BibitemOpen
  \bibfield  {author} {\bibinfo {author} {\bibfnamefont {T.}~\bibnamefont
  {Jacobson}},\ }\href {\doibase 10.1103/PhysRevLett.83.2699} {\bibfield
  {journal} {\bibinfo  {journal} {Phys. Rev. Lett.}\ }\textbf {\bibinfo
  {volume} {83}},\ \bibinfo {pages} {2699} (\bibinfo {year} {1999})},\ \Eprint
  {http://arxiv.org/abs/astro-ph/9905303} {arXiv:astro-ph/9905303 [astro-ph]}
  \BibitemShut {NoStop}%
\bibitem [{\citenamefont {Horbatsch}\ and\ \citenamefont
  {Burgess}(2012)}]{Horbatsch:2011ye}%
  \BibitemOpen
  \bibfield  {author} {\bibinfo {author} {\bibfnamefont {M.~W.}\ \bibnamefont
  {Horbatsch}}\ and\ \bibinfo {author} {\bibfnamefont {C.~P.}\ \bibnamefont
  {Burgess}},\ }\href {\doibase 10.1088/1475-7516/2012/05/010} {\bibfield
  {journal} {\bibinfo  {journal} {JCAP}\ }\textbf {\bibinfo {volume} {1205}},\
  \bibinfo {pages} {010} (\bibinfo {year} {2012})},\ \Eprint
  {http://arxiv.org/abs/1111.4009} {arXiv:1111.4009 [gr-qc]} \BibitemShut
  {NoStop}%
\bibitem [{\citenamefont {Freire}\ \emph {et~al.}(2012)\citenamefont {Freire},
  \citenamefont {Wex}, \citenamefont {Esposito-Farese}, \citenamefont
  {Verbiest}, \citenamefont {Bailes}, \citenamefont {Jacoby}, \citenamefont
  {Kramer}, \citenamefont {Stairs}, \citenamefont {Antoniadis},\ and\
  \citenamefont {Janssen}}]{Freire:2012mg}%
  \BibitemOpen
  \bibfield  {author} {\bibinfo {author} {\bibfnamefont {P.~C.~C.}\
  \bibnamefont {Freire}}, \bibinfo {author} {\bibfnamefont {N.}~\bibnamefont
  {Wex}}, \bibinfo {author} {\bibfnamefont {G.}~\bibnamefont
  {Esposito-Farese}}, \bibinfo {author} {\bibfnamefont {J.~P.~W.}\ \bibnamefont
  {Verbiest}}, \bibinfo {author} {\bibfnamefont {M.}~\bibnamefont {Bailes}},
  \bibinfo {author} {\bibfnamefont {B.~A.}\ \bibnamefont {Jacoby}}, \bibinfo
  {author} {\bibfnamefont {M.}~\bibnamefont {Kramer}}, \bibinfo {author}
  {\bibfnamefont {I.~H.}\ \bibnamefont {Stairs}}, \bibinfo {author}
  {\bibfnamefont {J.}~\bibnamefont {Antoniadis}}, \ and\ \bibinfo {author}
  {\bibfnamefont {G.~H.}\ \bibnamefont {Janssen}},\ }\href {\doibase
  10.1111/j.1365-2966.2012.21253.x} {\bibfield  {journal} {\bibinfo  {journal}
  {Mon. Not. Roy. Astron. Soc.}\ }\textbf {\bibinfo {volume} {423}},\ \bibinfo
  {pages} {3328} (\bibinfo {year} {2012})},\ \Eprint
  {http://arxiv.org/abs/1205.1450} {arXiv:1205.1450 [astro-ph.GA]} \BibitemShut
  {NoStop}%
\bibitem [{\citenamefont {Wex}(2014)}]{Wex:2014nva}%
  \BibitemOpen
  \bibfield  {author} {\bibinfo {author} {\bibfnamefont {N.}~\bibnamefont
  {Wex}},\ }\href@noop {} {\  (\bibinfo {year} {2014})},\ \Eprint
  {http://arxiv.org/abs/1402.5594} {arXiv:1402.5594 [gr-qc]} \BibitemShut
  {NoStop}%
\bibitem [{\citenamefont {Di~Casola}\ \emph {et~al.}(2015)\citenamefont
  {Di~Casola}, \citenamefont {Liberati},\ and\ \citenamefont
  {Sonego}}]{DiCasola:2013iia}%
  \BibitemOpen
  \bibfield  {author} {\bibinfo {author} {\bibfnamefont {E.}~\bibnamefont
  {Di~Casola}}, \bibinfo {author} {\bibfnamefont {S.}~\bibnamefont {Liberati}},
  \ and\ \bibinfo {author} {\bibfnamefont {S.}~\bibnamefont {Sonego}},\ }\href
  {\doibase 10.1119/1.4895342} {\bibfield  {journal} {\bibinfo  {journal} {Am.
  J. Phys.}\ }\textbf {\bibinfo {volume} {83}},\ \bibinfo {pages} {39}
  (\bibinfo {year} {2015})},\ \Eprint {http://arxiv.org/abs/1310.7426}
  {arXiv:1310.7426 [gr-qc]} \BibitemShut {NoStop}%
\bibitem [{\citenamefont {Bertotti}\ and\ \citenamefont
  {Grishchuk}(1990)}]{0264-9381-7-10-007}%
  \BibitemOpen
  \bibfield  {author} {\bibinfo {author} {\bibfnamefont {B.}~\bibnamefont
  {Bertotti}}\ and\ \bibinfo {author} {\bibfnamefont {L.~P.}\ \bibnamefont
  {Grishchuk}},\ }\href {http://stacks.iop.org/0264-9381/7/i=10/a=007}
  {\bibfield  {journal} {\bibinfo  {journal} {Classical and Quantum Gravity}\
  }\textbf {\bibinfo {volume} {7}},\ \bibinfo {pages} {1733} (\bibinfo {year}
  {1990})}\BibitemShut {NoStop}%
\bibitem [{\citenamefont {Uzan}(2011)}]{uzan:2010pm}%
  \BibitemOpen
  \bibfield  {author} {\bibinfo {author} {\bibfnamefont {J.-P.}\ \bibnamefont
  {Uzan}},\ }\href {\doibase 10.12942/lrr-2011-2} {\bibfield  {journal}
  {\bibinfo  {journal} {Living Rev. Rel.}\ }\textbf {\bibinfo {volume} {14}},\
  \bibinfo {pages} {2} (\bibinfo {year} {2011})},\ \Eprint
  {http://arxiv.org/abs/1009.5514} {arXiv:1009.5514 [astro-ph.CO]} \BibitemShut
  {NoStop}%
\bibitem [{\citenamefont {Will}(2006)}]{Will2006}%
  \BibitemOpen
  \bibfield  {author} {\bibinfo {author} {\bibfnamefont {C.~M.}\ \bibnamefont
  {Will}},\ }\href {\doibase 10.12942/lrr-2006-3} {\bibfield  {journal}
  {\bibinfo  {journal} {Living Reviews in Relativity}\ }\textbf {\bibinfo
  {volume} {9}},\ \bibinfo {pages} {3} (\bibinfo {year} {2006})}\BibitemShut
  {NoStop}%
\bibitem [{\citenamefont {Yagi}\ \emph {et~al.}(2014)\citenamefont {Yagi},
  \citenamefont {Blas}, \citenamefont {Barausse},\ and\ \citenamefont
  {Yunes}}]{Yagi:2013ava}%
  \BibitemOpen
  \bibfield  {author} {\bibinfo {author} {\bibfnamefont {K.}~\bibnamefont
  {Yagi}}, \bibinfo {author} {\bibfnamefont {D.}~\bibnamefont {Blas}}, \bibinfo
  {author} {\bibfnamefont {E.}~\bibnamefont {Barausse}}, \ and\ \bibinfo
  {author} {\bibfnamefont {N.}~\bibnamefont {Yunes}},\ }\href {\doibase
  10.1103/PhysRevD.90.069902, 10.1103/PhysRevD.90.069901,
  10.1103/PhysRevD.89.084067} {\bibfield  {journal} {\bibinfo  {journal} {Phys.
  Rev.}\ }\textbf {\bibinfo {volume} {D89}},\ \bibinfo {pages} {084067}
  (\bibinfo {year} {2014})},\ \bibinfo {note} {[Erratum: Phys.
  Rev.D90,no.6,069901(2014)]},\ \Eprint {http://arxiv.org/abs/1311.7144}
  {arXiv:1311.7144 [gr-qc]} \BibitemShut {NoStop}%
\bibitem [{\citenamefont {Seoane}\ \emph {et~al.}(2013)\citenamefont {Seoane}
  \emph {et~al.}}]{Seoane:2013qna}%
  \BibitemOpen
  \bibfield  {author} {\bibinfo {author} {\bibfnamefont {P.~A.}\ \bibnamefont
  {Seoane}} \emph {et~al.} (\bibinfo {collaboration} {eLISA}),\ }\href@noop {}
  {\  (\bibinfo {year} {2013})},\ \Eprint {http://arxiv.org/abs/1305.5720}
  {arXiv:1305.5720 [astro-ph.CO]} \BibitemShut {NoStop}%
\bibitem [{\citenamefont {Audley}\ \emph {et~al.}(2017)\citenamefont {Audley}
  \emph {et~al.}}]{Audley:2017drz}%
  \BibitemOpen
  \bibfield  {author} {\bibinfo {author} {\bibfnamefont {H.}~\bibnamefont
  {Audley}} \emph {et~al.} (\bibinfo {collaboration} {LISA}),\ }\href@noop {}
  {\  (\bibinfo {year} {2017})},\ \Eprint {http://arxiv.org/abs/1702.00786}
  {arXiv:1702.00786 [astro-ph.IM]} \BibitemShut {NoStop}%
\bibitem [{\citenamefont {Chamberlain}\ and\ \citenamefont
  {Yunes}(2017)}]{Chamberlain:2017fjl}%
  \BibitemOpen
  \bibfield  {author} {\bibinfo {author} {\bibfnamefont {K.}~\bibnamefont
  {Chamberlain}}\ and\ \bibinfo {author} {\bibfnamefont {N.}~\bibnamefont
  {Yunes}},\ }\href {\doibase 10.1103/PhysRevD.96.084039} {\bibfield  {journal}
  {\bibinfo  {journal} {Phys. Rev.}\ }\textbf {\bibinfo {volume} {D96}},\
  \bibinfo {pages} {084039} (\bibinfo {year} {2017})},\ \Eprint
  {http://arxiv.org/abs/1704.08268} {arXiv:1704.08268 [gr-qc]} \BibitemShut
  {NoStop}%
\bibitem [{\citenamefont {Carson}\ \emph {et~al.}(2019)\citenamefont {Carson},
  \citenamefont {Seymour},\ and\ \citenamefont {Yagi}}]{Carson:2019fxr}%
  \BibitemOpen
  \bibfield  {author} {\bibinfo {author} {\bibfnamefont {Z.}~\bibnamefont
  {Carson}}, \bibinfo {author} {\bibfnamefont {B.~C.}\ \bibnamefont {Seymour}},
  \ and\ \bibinfo {author} {\bibfnamefont {K.}~\bibnamefont {Yagi}},\
  }\href@noop {} {\  (\bibinfo {year} {2019})},\ \Eprint
  {http://arxiv.org/abs/1907.03897} {arXiv:1907.03897 [gr-qc]} \BibitemShut
  {NoStop}%
\bibitem [{\citenamefont {Carson}\ and\ \citenamefont
  {Yagi}()}]{carsonyagi:2019}%
  \BibitemOpen
  \bibfield  {author} {\bibinfo {author} {\bibfnamefont {Z.}~\bibnamefont
  {Carson}}\ and\ \bibinfo {author} {\bibfnamefont {K.}~\bibnamefont {Yagi}},\
  }\href@noop {} {\bibinfo  {journal} {In preparation}\ }\BibitemShut {NoStop}%
\bibitem [{\citenamefont {Khan}\ \emph {et~al.}(2016)\citenamefont {Khan},
  \citenamefont {Husa}, \citenamefont {Hannam}, \citenamefont {Ohme},
  \citenamefont {Pürrer}, \citenamefont {Jiménez~Forteza},\ and\
  \citenamefont {Bohé}}]{Khan:2015jqa}%
  \BibitemOpen
\bibfield  {journal} {  }\bibfield  {author} {\bibinfo {author} {\bibfnamefont
  {S.}~\bibnamefont {Khan}}, \bibinfo {author} {\bibfnamefont {S.}~\bibnamefont
  {Husa}}, \bibinfo {author} {\bibfnamefont {M.}~\bibnamefont {Hannam}},
  \bibinfo {author} {\bibfnamefont {F.}~\bibnamefont {Ohme}}, \bibinfo {author}
  {\bibfnamefont {M.}~\bibnamefont {Pürrer}}, \bibinfo {author} {\bibfnamefont
  {X.}~\bibnamefont {Jiménez~Forteza}}, \ and\ \bibinfo {author}
  {\bibfnamefont {A.}~\bibnamefont {Bohé}},\ }\href {\doibase
  10.1103/PhysRevD.93.044007} {\bibfield  {journal} {\bibinfo  {journal} {Phys.
  Rev.}\ }\textbf {\bibinfo {volume} {D93}},\ \bibinfo {pages} {044007}
  (\bibinfo {year} {2016})},\ \Eprint {http://arxiv.org/abs/1508.07253}
  {arXiv:1508.07253 [gr-qc]} \BibitemShut {NoStop}%
\end{thebibliography}%
\end{document}